\documentclass{article}

\PassOptionsToPackage{numbers, compress}{natbib}

\usepackage[final]{templatepackage/neurips_2024}

\usepackage[utf8]{inputenc} \usepackage[T1]{fontenc}    \usepackage{hyperref}       \usepackage{url}            \usepackage{booktabs}       \usepackage{amsfonts}       \usepackage{nicefrac}       \usepackage{microtype}

\usepackage{physics}
\usepackage{amsfonts}
\ifcsname proof\endcsname
\else
\usepackage{amsthm}
\fi
\usepackage{amssymb}
\usepackage{cleveref}
\usepackage{hhline}
\usepackage{multirow}
\usepackage{enumitem}
\ifcsname citet\endcsname
\else
  \usepackage{natbib}
\fi
\ifcsname cellcolor\endcsname
\else
  \usepackage[table]{xcolor}
\fi
\ifcsname mathclap\endcsname
\else
  \usepackage{mathtools}
\fi
\ifcsname tcolorbox\endcsname
\else
  \usepackage{tcolorbox}
\fi
\usepackage{subcaption}
\usepackage{bm}
\usepackage{xfrac} \newcommand{\opn}[1]{\operatorname{#1}}
\DeclarePairedDelimiterXPP\inner[2]{}{\langle}{\rangle}{}{#1,#2}
\allowdisplaybreaks
\makeatletter
\newcommand{\myvspace}{\@ifstar\myvspacestar\myvspacenostar}
\newcommand{\myvspacenostar}[1]{}
\newcommand{\myvspacestar}[1]{}
\makeatother \ifcsname theorem\endcsname
\else
    \newtheorem{theorem}{Theorem}
\fi
\ifcsname definition\endcsname
\else
    
\fi
\ifcsname lemma\endcsname
\else
    \newtheorem{lemma}[theorem]{Lemma}
\fi
\ifcsname remark\endcsname
\else
    \newtheorem{remark}[theorem]{Remark}
\fi
 \ifcsname algorithmic\endcsname
\else
  \usepackage{algorithm}
  \usepackage{algorithmic}
\fi

  \usepackage{tikz}
\usetikzlibrary{arrows.meta, positioning}
\usetikzlibrary{calc,math}  
\title{Inevitable Trade-off between Watermark Strength and Speculative Sampling Efficiency for Language Models}

\author{Zhengmian Hu, Heng Huang \\
  Department of Computer Science\\
  University of Maryland\\
  College Park, MD 20742 \\
  \texttt{huzhengmian@gmail.com,heng@umd.edu} \\
}

\begin{document}

\maketitle

\begin{abstract}
Large language models are probabilistic models, and the process of generating content is essentially sampling from the output distribution of the language model. Existing watermarking techniques inject watermarks into the generated content without altering the output quality. On the other hand, existing acceleration techniques, specifically speculative sampling, leverage a draft model to speed up the sampling process while preserving the output distribution. However, there is no known method to simultaneously accelerate the sampling process and inject watermarks into the generated content. In this paper, we investigate this direction and find that the integration of watermarking and acceleration is non-trivial. We prove a no-go theorem, which states that it is impossible to simultaneously maintain the highest watermark strength and the highest sampling efficiency. Furthermore, we propose two methods that maintain either the sampling efficiency or the watermark strength, but not both. Our work provides a rigorous theoretical foundation for understanding the inherent trade-off between watermark strength and sampling efficiency in accelerating the generation of watermarked tokens for large language models. We also conduct numerical experiments to validate our theoretical findings and demonstrate the effectiveness of the proposed methods.
\end{abstract} \section{Introduction}\label{se:introduction}
Large language models (LLMs) have demonstrated remarkable performance in various natural language processing tasks, enabling a wide range of applications such as chatbots \citep{luo2022critical}, content generation \citep{li2024pre}, code generation \citep{chen2021evaluating}, and more. However, the high training and inference costs of LLMs pose significant challenges. The substantial computational resources along with the high latency during inference can negatively impact user experience and limit their potential applications.

To address the issue of high inference costs, speculative sampling \citep{leviathan2023fast,chen2023accelerating} has emerged as a promising approach. This technique leverages a smaller, faster draft model to generate candidate results, which are then validated and corrected by a larger, more accurate target model. Compared with other acceleration methods such as knowledge distillation, model quantization, and model pruning, the key advantage of speculative sampling is that it can significantly reduce inference latency without compromising the quality of the generated content.

In addition to the challenge of high inference costs, protecting the intellectual property rights of LLMs generated content has become increasingly important. Digital watermarking techniques \citep{aaron2022blog,kirchenbauer2023watermark} have been proposed to embed watermark information into the generated content, enabling the tracking of model usage. Unbiased watermarking schemes \citep{hu2023unbiased} have been developed to ensure that the watermarking process does not affect the quality of the generated content.

A natural question arises: can we leverage speculative sampling to accelerate the generation of watermarked content? To address this question, we propose a general framework called the \textit{two reweight framework}, which allows for the integration of unbiased watermarking and speculative sampling techniques while guaranteeing an unchanged output distribution. The main innovation of our framework lies in the simultaneous reweighting of both the target model and the draft model, which improves the sampling efficiency compared to naively applying speculative sampling to a watermarked target model.

To evaluate the effectiveness of our framework, we consider two key metrics: watermark strength and acceleration performance. A fundamental question is whether it is possible to achieve both strong watermarking and efficient speculative sampling simultaneously. Specifically, we aim to answer the following question: 
\begin{center}
\begin{tcolorbox}\centering
\textit{Can we obtain the same watermark strength as in the case without acceleration while maintaining the same sampling efficiency as in the case without watermarking?
}
\end{tcolorbox}
\end{center}

Surprisingly, we got a negative answer to this question. We prove a no-go theorem, which states that under the \textit{two reweight framework}, it is impossible to simultaneously maintain both the watermark strength and the sampling efficiency when the vocabulary size is greater than two. This result highlights the inherent trade-off between watermarking and acceleration in the context of large language models.

To better explore the trade-offs between these two objectives, we propose two practical algorithms within the \textit{two reweight framework}. The first algorithm focuses on maintaining the watermark strength, while the second algorithm aims to maintain the sampling efficiency.

The main contributions of this paper are as follows:
\begin{itemize}[leftmargin=0.15in]
    \item We propose the \textit{two reweight framework}, a general framework that allows for the integration of unbiased watermarking and speculative sampling techniques while ensuring an unchanged output distribution.
    \item We prove a no-go theorem, which states that under the \textit{two reweight framework}, it is impossible to simultaneously maintain both the watermark strength and the sampling efficiency when the vocabulary size is greater than two.
    \item We propose two practical algorithms within the \textit{two reweight framework} that focus on maintaining either the watermark strength or the sampling efficiency, providing insights into the achievable trade-offs.
\end{itemize}

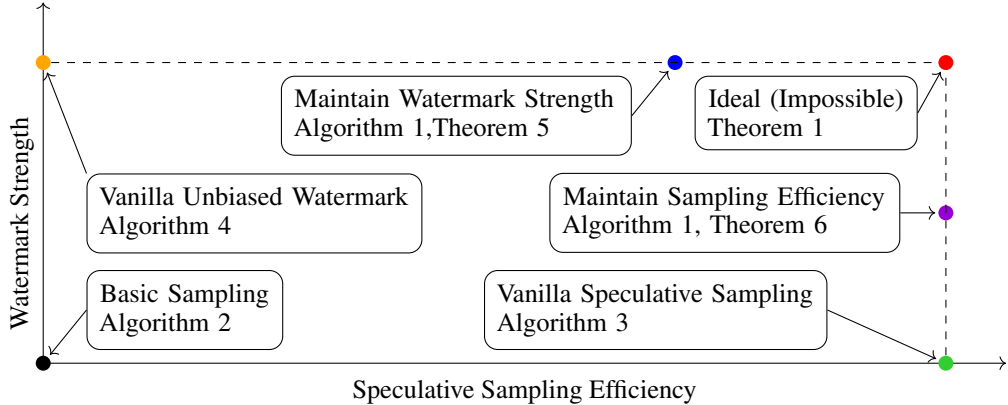
\begin{figure}[t]
\myvspace{-20pt}
\centering
\begin{tikzpicture}[scale=0.8]
\definecolor{basic}{RGB}{0,0,0}
\definecolor{vanilla_watermark}{RGB}{255,165,0}
\definecolor{vanilla_sampling}{RGB}{50,205,50}
\definecolor{ideal}{RGB}{255,0,0}
\definecolor{maintain_watermark}{RGB}{0,0,255}
\definecolor{maintain_sampling}{RGB}{148,0,211}
\definecolor{block_color}{RGB}{255,255,255}

\tikzmath{
 \WIDTH=16; \HEIGHT=6; \WM=5; \SE=15;
};

\draw [<->] (0,\HEIGHT) -- (0,0) -- (\WIDTH,0);
\node[rotate=90, above=0.5cm] at (0,\HEIGHT/2-1.3) {Watermark Strength};
\node[below=0.1cm] at (\WIDTH/2,0) {Speculative Sampling Efficiency};

\node[circle,fill=basic,inner sep=2pt] (basic) at (0,0) {};
\node[circle,fill=vanilla_watermark,inner sep=2pt] (vanilla_watermark) at (0,\WM) {};
\node[circle,fill=vanilla_sampling,inner sep=2pt] (vanilla_sampling) at (\SE,0) {};
\node[circle,fill=ideal,inner sep=2pt] (ideal) at (\SE,\WM) {};
\node[circle,fill=maintain_watermark,inner sep=2pt] (maintain_watermark) at (7*\SE/10,\WM) {};
\node[circle,fill=maintain_sampling,inner sep=2pt] (maintain_sampling) at (\SE,\WM/2) {};

\draw[dashed] (vanilla_watermark) -- (ideal);
\draw[dashed] (vanilla_sampling) -- (ideal);

\node[above right=0.1cm and 0.5cm of basic, text width=2.25cm, align=left, fill=block_color, rounded corners=5pt, inner sep=5pt, draw, thin] (text_basic) {Basic Sampling\\ \Cref{alg:basic}};
\draw[->,shorten >=1pt] (text_basic.west) -- (basic);

\node[below right=1.4cm and 0.5cm of vanilla_watermark, text width=4.1cm, align=left, fill=block_color, rounded corners=5pt, inner sep=5pt, draw, thin] (text_vanilla_watermark) {Vanilla Unbiased Watermark\\ \Cref{alg:wm}};
\draw[->,shorten >=1pt] (text_vanilla_watermark.north west) -- (vanilla_watermark);

\node[above left=0.1cm and 1.5cm of vanilla_sampling, text width=4.2cm, align=left, fill=block_color, rounded corners=5pt, inner sep=5pt, draw, thin] (text_vanilla_sampling) {Vanilla Speculative Sampling\\ \Cref{alg:sps}};
\draw[->,shorten >=1pt] (text_vanilla_sampling.east) -- (vanilla_sampling);

\node[below left=0.1cm and 0.3cm of ideal, text width=2.6cm, align=left, fill=block_color, rounded corners=5pt, inner sep=5pt, draw, thin] (text_ideal) {Ideal (Impossible)\\ \Cref{thm:no_go}};
\draw[->,shorten >=1pt] (text_ideal.east) -- (ideal);

\node[below left=0.1cm and 0.5cm of maintain_watermark, text width=4.3cm, align=left, fill=block_color, rounded corners=5pt, inner sep=5pt, draw, thin] (text_maintain_watermark) {Maintain Watermark Strength\\ \Cref{alg:both_algorithm},\Cref{thm:maintain_watermark}};
\draw[->,shorten >=1pt] (text_maintain_watermark.east) -- (maintain_watermark);

\node[left=0.5cm of maintain_sampling, text width=4.3cm, align=left, fill=block_color, rounded corners=5pt, inner sep=5pt, draw, thin] (text_maintain_sampling) {Maintain Sampling Efficiency\\ \Cref{alg:both_algorithm}, \Cref{thm:maintain_efficiency}};
\draw[->,shorten >=1pt] (text_maintain_sampling) -- (maintain_sampling);

\end{tikzpicture}
\caption{Taxonomy of watermarking and speculative sampling trade-offs in language models. The ideal case of maintaining both watermark strength and sampling efficiency is proven to be impossible by the no-go theorem. The proposed algorithms focus on maintaining either watermark strength or sampling efficiency.}
\label{fig:contributions}
\myvspace{-10pt}
\end{figure} 
To the best of our knowledge, this work represents the first exploration of the intersection between unbiased watermarking and speculative sampling, introducing a novel framework, a significant no-go theorem, and pioneering practical algorithms.

 \myvspace{-10pt}
\section{Preliminary}\label{se:preliminary}
\myvspace{-10pt}

In this section, we will introduce the basic concepts and notations used throughout the paper, and provide a brief overview of watermarking and speculative sampling techniques for large language models.

A language model defines a probability distribution over sequences of tokens from a vocabulary set $\Sigma$. It assigns a probability $P(x_{n+1}|x_1,x_2,...,x_n)$ to the next token $x_{n+1}$ given the context of previous tokens $x_1,x_2,...,x_n$. We use $\Delta_\Sigma$ to denote the set of all possible probability distributions over the vocabulary $\Sigma$.

Following \citet{hu2023unbiased}, we define a watermarking scheme as a tuple $(\mathcal{E},P_E,R)$, where $\mathcal{E}$ is a set of watermark codes, $P_E$ is a probability distribution over $\mathcal{E}$, and $R:\mathcal{E}\times\Delta_\Sigma\to\Delta_\Sigma$ is a reweighting function that maps a watermark code $E\in\mathcal{E}$ and a probability distribution $P\in\Delta_\Sigma$ to a watermarked distribution $R_E(P)\in\Delta_\Sigma$. We focus on unbiased watermarking schemes that satisfy $\mathbb{E}_{E\sim P_E}[R_E(P)]=P$ for all $P\in\Delta_\Sigma$, unless explicitly stated otherwise.

To generate a watermarked token $x$, we first compute a watermark code $E\sim P_E$ based on the context, and then sample the token from the watermarked distribution, i.e., $x\sim R_E(P)$. 
The entropy of the distribution $P$ determines the maximum amount of watermark that can be injected. For a distribution $P$ with high entropy, the divergence between the watermarked distribution $R_E(P)$ and the original distribution $P$ can be larger, allowing for more watermark information to be injected.

The presence of the watermark can be detected by statistical tests. The pivotal quantity used in these tests is often referred to as the watermark score. A higher watermark score implies a more detectable watermark. The log likelihood ratio (LLR) is the most powerful score for detecting watermarks in the absence of any perturbations. However, in practice, more robust scores such as the maximin-LLR or likelihood-agnostic scores are often used.
In this paper, we consider two specific watermark scores: the maximin-LLR score, which is described in detail in \citep{hu2023unbiased}, and the U score, which is a likelihood-agnostic score that can be defined for both DeltaGumbel reweight and Gamma reweight schemes. The details of the U score, DeltaGumbel reweight and Gamma reweight are provided in \Cref{se:Uscore}.

The P-value can be computed by considering the absence of a watermark as the null hypothesis. For a score $S$ with a known moment-generating function (MGF), the P-value can be upper bounded using the Chernoff bound:
\myvspace{-4pt}
\begin{equation}\label{eq:p_value}
P_{\text{null}}(S\geq\hat{S})\leq\min_{\lambda\geq0}\mathbb{E}[e^{\lambda S}]\exp(-\lambda\hat{S}).
\end{equation}
\par\myvspace{-4pt}

Speculative sampling \citep{leviathan2023fast,chen2023accelerating} is a technique for accelerating the generation of tokens from a target model $P$ by leveraging a faster draft model $Q$. The key idea is to first sample a draft token $\tilde{x}$ from the draft model $Q$, and then accept or reject it based on the ratio of the target and draft probabilities. If the draft token is rejected, a new token is sampled from a residual distribution proportional to the difference between the target and draft probabilities. Formally, the speculative sampling process generates a token $x$ as follows:
\begin{equation}
    \mathcal{P}(x=j|\tilde{x}=i)=\begin{cases}\min(1,\frac{P(i)}{Q(i)})&\text{if }i=j,\\\frac{(1-\frac{P(i)}{Q(i)})_+(P(j)-Q(j))_+}{\sum_{z\in\Sigma}(Q(z)-P(z))_+}&\text{if }i\ne j,\end{cases}
\end{equation}
where $(x)_+=\max(0,x)$. The design of the speculative process ensures that the final distribution of the generated token $x$ matches the target distribution $P$. The efficiency of speculative sampling can be measured by the overlap probability $\alpha(P,Q)=\sum_{t\in\Sigma}\min(P(t),Q(t))$, which is the probability of accepting the draft token in each step. The overlap probability is related to the total variation distance between $P$ and $Q$ by $\opn{TV}(P,Q)=1-\alpha(P,Q)$. Such a speculative sampling process can be applied multiple times to generate and verify multiple draft tokens in one step.

Due to space limitations, we have moved the discussion of other related works to the \Cref{se:related}. \myvspace{-4pt}
\section{Two Reweight Framework for Accelerated Generation of Watermarked Tokens}\label{se:tworeweight}
\myvspace{-4pt}

In this section, we propose a novel framework called the \textit{two reweight framework} for accelerating the generation of watermarked tokens based on speculative sampling techniques. The motivation behind this non-trivial framework is that naively applying speculative sampling to a watermarked target distribution $R_E(P)$ may significantly reduce the overlap probability $\alpha(R_E(P),Q)$ with the draft distribution $Q$, leading to a small sampling efficiency.

The key innovation of the \textit{two reweight framework} is to apply a separate reweighting function $R'$ to the draft distribution $Q$, using the same watermark code $E$ as the one used for reweighting the target distribution. By doing so, we aim to increase the overlap probability between the watermarked target distribution $R_E(P)$ and the watermarked draft distribution $R'_E(Q)$, i.e., $\alpha(R_E(P),R'_E(Q))$, thus improving the sampling efficiency.

Formally, we define the watermarked draft distribution using another reweighting function $(\mathcal{E},P_E,R')$, where $R':\mathcal{E}\times\Delta_\Sigma\to\Delta_\Sigma$ is a function that maps a watermark code $E\in\mathcal{E}$ and a draft distribution $Q\in\Delta_\Sigma$ to a watermarked draft distribution $R'_E(Q)\in\Delta_\Sigma$. The framework itself does not require the watermarked draft distribution to be unbiased, i.e., $\mathbb{E}_{E\sim P_E}[R'_E(Q)]=Q$ for all $Q\in\Delta_\Sigma$. However, we will see later that this unbiasedness property naturally emerges when we require the final output distribution to be unbiased and aim to improve the sampling efficiency  (\Cref{lem:unbiased_draft}).

To generate a watermarked token, we first sample a draft token $\tilde{x}$ from the watermarked draft distribution, i.e., $\tilde{x}\sim R'_E(Q)$ or equivalently $\mathcal{P}(\tilde{x}=i)=R'_E(Q)(i)$ for all $i\in\Sigma$. Then, we perform certain speculative sampling based on the draft token to obtain the generated token $x$. The speculative process is defined by a conditional probability distribution $A(j|i)$ for all $i,j\in\Sigma$, where $A(\cdot|i)\in\Delta_\Sigma$ for each $i$. The design of $A$ can depend on the target distribution $P$, the draft distribution $Q$, and the watermark code $E$. The probability of generating a token $x=j$ given a draft token $\tilde{x}=i$ is given by $\mathcal{P}(x=j|\tilde{x}=i)=A(j|i)$.

The distribution of the generated token, which we call the generation distribution, can be computed as follows:
\begin{equation}
    \mathcal{P}(x=j)=\sum_{i\in\Sigma} \mathcal{P}(x=j|\tilde{x}=i) \mathcal{P}(\tilde{x}=i)=\sum_{i\in\Sigma} A(j|i)R'_E(Q)(i)=(A\circ R'_E(Q))(j).
\end{equation}
We denote the generation distribution by $\widehat{R}_E(P)=A\circ R'_E(Q)$.

To ensure that the \textit{two reweight framework} produces an unbiased output distribution, we require that for all $P\in\Delta_\Sigma$:
\myvspace{-4pt}
\begin{equation}
    \mathbb{E}_{E\sim P_E}[\widehat{R}_E(P)]=P.
\end{equation}
\par\myvspace{-4pt} \myvspace{-10pt}
\section{No-go Theorem}\label{se:no_go}
\myvspace{-4pt}

Despite the potential of the \textit{two reweight framework}, we present a no-go theorem that shows the impossibility of simultaneously maintaining the watermark strength and sampling efficiency when the vocabulary size is greater than two.

\begin{theorem}[No-go Theorem]\label{thm:no_go}
When the vocabulary size $\abs{\Sigma}>2$, there do not exist non-trivial reweighting functions $R:\mathcal{E}\times\Delta_\Sigma\to\Delta_\Sigma$ and $R':\mathcal{E}\times\Delta_\Sigma\to\Delta_\Sigma$, and a speculative process $A(j|i)$ such that for all $P,Q\in\Delta_\Sigma$:
\myvspace{-5pt}
\begin{enumerate}
    \item The watermark strength is maintained: $\widehat{R}_E(P)=R_E(P)$.
    \item The sampling efficiency is maintained: $\alpha(P,Q)=\mathbb{E}_{E\sim P_E} [\sum_{i\in\Sigma}A(i|i)R'_E(Q)(i)]$.
\end{enumerate}
\end{theorem}

\begin{remark}[Condition for maintaining the watermark strength]
The condition $\widehat{R}_E(P)=R_E(P)$ in \Cref{thm:no_go} ensures that the watermark strength is maintained by keeping the average watermark score unchanged, i.e.,
\begin{equation}\label{eq:watermark_score}
    \underbrace{\mathbb{E}_{E\sim P_E} \mathbb{E}_{t\sim R_E(P)}[Score(t,E)]}_{w:=}=\underbrace{\mathbb{E}_{E\sim P_E} \mathbb{E}_{t\sim \widehat{R}_E(P)}[Score(t,E)]}_{w':=},
\end{equation}
where $Score(t,E)$ is an arbitrary function that measures the watermark strength.

Strictly speaking, to ensure that the watermark strength remains unchanged, we only need to require $w=w'$, and the condition $\widehat{R}_E(P)=R_E(P)$ is a sufficient condition. However, due to the large design space of scoring functions, if we want to maintain $w=w'$ for every possible score, then $\widehat{R}_E(P)=R_E(P)$ becomes a necessary condition.

On the other hand, for a fixed scoring function and a specific $R_E$, it is possible that $\widehat{R}_E(P)\neq R_E(P)$ while $w=w'$ or even $w<w'$. In other words, the condition $\widehat{R}_E(P)=R_E(P)$ is not always necessary for maintaining the watermark strength for a specific scoring function and reweighting function.
\end{remark}

The proof of the no-go theorem relies on the following two lemmas, which reveal the connections between maintaining the sampling efficiency, maintaining the watermark strength, and the properties of the reweighting functions.

\begin{lemma}[Maintaining Sampling Efficiency Implies Unbiased Watermarked Draft Model]\label{lem:unbiased_draft}
If for all $P,Q\in\Delta_\Sigma$, we have
\begin{equation*}
    \alpha(P,Q)=\mathbb{E}_{E\sim P_E} \left[\sum_{i\in\Sigma}A(i|i)R'_E(Q)(i)\right],
\end{equation*}
\par\myvspace{-10pt}
then $\mathbb{E}_{E\sim P_E}[R'_E(Q)]=Q$ for all $Q\in\Delta_\Sigma$.
\end{lemma} \begin{lemma}[Maintaining Watermark Strength and Sampling Efficiency Implies Same Reweight Function]\label{lem:same_reweight}
Under the \textit{two reweight framework}, if for all $P,Q\in\Delta_\Sigma$, we have
\begin{equation*}
    \alpha(P,Q)=\mathbb{E}_{E\sim P_E} \left[\sum_{i\in\Sigma}A(i|i)R'_E(Q)(i)\right],
    \quad\quad
    \widehat{R}_E(P)=R_E(P),
\end{equation*}
\par\myvspace{-10pt}
then $R'_E(Q)=R_E(Q)$ for all $Q\in\Delta_\Sigma$.
\end{lemma} 
The proofs of these lemmas are deferred to \Cref{se:proofs}. With these lemmas, we can now prove the no-go theorem.

\begin{proof}[Proof of \Cref{thm:no_go}]
According to \Cref{lem:same_reweight}, maintaining both the watermark strength and sampling efficiency under the \textit{two reweight framework} implies that $R'_E(Q)=R_E(Q)$ for all $Q\in\Delta_\Sigma$. Therefore, we have
\begin{equation}\label{eq:alpha_inequality}
    \alpha(P,Q)\leq\mathbb{E}_{E\sim P_E} [\alpha(R_E(P),R_E(Q))].
\end{equation}
To see this, note that
\myvspace{-10pt}
\begin{gather*}
    R_E(P)(i)=\widehat{R}_E(P)(i)=\sum_j A(i|j) R_E(Q)(j)\geq A(i|i) R_E(Q)(i),\\
    R_E(Q)(i)\geq A(i|i) R_E(Q)(i),\\
    A(i|i) R_E(Q)(i)\leq\min(R_E(Q)(i),R_E(P)(i)).
\end{gather*}
\par\myvspace{-10pt}
Summing over $i$, we get
\begin{equation*}
    \sum_i A(i|i) R_E(Q)(i)\leq\sum_i\min(R_E(Q)(i),R_E(P)(i))=\alpha(R_E(Q),R_E(P)).
\end{equation*}
\par\myvspace{-10pt}
Taking the expectation over $E$, we obtain \Cref{eq:alpha_inequality}.

Recall that $\alpha(P,Q)=1-\opn{TV}(P,Q)$, where $\opn{TV}(P,Q)$ denotes the total variation distance between $P$ and $Q$. Therefore, \Cref{eq:alpha_inequality} is equivalent to
\myvspace{-4pt}
\begin{equation}\label{eq:tv_equality}
    \opn{TV}(P,Q)\geq\mathbb{E}_{E\sim P_E} [\opn{TV}(R_E(P),R_E(Q))].
\end{equation}
\par\myvspace{-10pt}

Viewing $P,Q\in\Delta_\Sigma$ as $n$-dimensional vectors, where $n=\abs{\Sigma}$, we can express the total variation distance as
\myvspace{-10pt}
\begin{equation}\label{eq:tv_max}
    2\opn{TV}(P,Q)=\max_{u\in[-1,1]^n} \inner{u}{P-Q},
\end{equation}
\par\myvspace{-4pt}
where the maximum is attained at $u^\ast(P,Q) = \opn{sign}(P-Q)$. Using this expression, we have
\myvspace{-4pt}
\begin{align}
    \mathbb{E}_{E\sim P_E} 2\opn{TV}(R_E(P),R_E(Q))&=\mathbb{E}_{E\sim P_E} \left[\max_{u\in[-1,1]^n} \inner{u}{R_E(P)-R_E(Q)}\right]\nonumber\\
    &\geq\mathbb{E}_{E\sim P_E}\inner{u^\ast(P,Q)}{R_E(P)-R_E(Q)}\label{eq:tv_inequality}\\
    &=\inner{u^\ast(P,Q)}{P-Q}=2\opn{TV}(P,Q).\label{eq:tv_inequality_final}
\end{align}
Combining \Cref{eq:tv_equality,eq:tv_inequality_final}, we conclude that the equality in \Cref{eq:tv_inequality} must hold, which is equivalent to
\begin{equation}\label{eq:u_star_condition}
    u^\ast(P,Q)\in\opn{Argmax}\limits_{u\in[-1,1]^n} \inner{u}{R_E(P)-R_E(Q)},
\end{equation}
almost surely for random $E$. This condition is equivalent to the following:
\begin{align}
    (P-Q)(i)=0&\implies(R_E(P)-R_E(Q))(i)=0,\label{eq:condition_1}\\
    (P-Q)(i)\geq0&\implies(R_E(P)-R_E(Q))(i)\geq0,\label{eq:condition_2}\\
    (P-Q)(i)\leq0&\implies(R_E(P)-R_E(Q))(i)\leq0,\label{eq:condition_3}
\end{align}
almost surely for random $E$ and for all $i\in\Sigma$.

Now, let us label the symbols in the vocabulary $\Sigma$ as $i\in\{1,\dots,n\}$. For a distribution $P=(p_1,p_2,\dots,p_n)$, define
\begin{equation*}
    T_i(j)=\begin{cases}
        p_i&j=i,\\
        1-p_i&j=i \opn{mod} n+1,\\
        0&\text{otherwise}.
    \end{cases}
\end{equation*}
For example, $T_1=(p_1,1-p_1,0,\dots,0)$, $T_2=(0,p_2,1-p_2,0,\dots,0)$, and $T_n=(1-p_n,0,\dots,0,p_n)$. Let functions $F_i(p_i)=R_E(T_i)(i)$. We claim that
\begin{equation}\label{eq:R_E_P_i}
    R_E(P)(i)=F_i(p_i).
\end{equation}
To see this, note that $(P-T_i)(i)=p_i-p_i=0$, so by \Cref{eq:condition_1}, we have $(R_E(P)-R_E(T_i))(i)=0$, which implies $R_E(P)(i)=R_E(T_i)(i)=F_i(p_i)$ almost surely.

Next, we show that the functions $F_i$ satisfy the following properties:
\begin{align}
    &F_i(0)=0,\label{eq:F_i_0}\\
    &F_i(1)=1,\label{eq:F_i_1}\\
    &F_i(p)\text{ is monotonically increasing in }p,\label{eq:F_i_monotone}\\
    &\sum_i p_i=1\implies \sum_i F_i(p_i)=1.\label{eq:F_i_sum}
\end{align}
\par\myvspace{-10pt}

To prove \Cref{eq:F_i_0}, consider the case when $p_i=0$. In this case, $T_i(j)=1$ if $j=i \opn{mod} n+1$ and $T_i(j)=0$ otherwise. To ensure the unbiasedness of the reweighting function, we must have $\mathbb{E}_{E\sim P_E}[R_E(T_i)]=T_i$, which implies $R_E(T_i)=T_i$ almost surely. Therefore, $R_E(T_i)(i)=T_i(i)=p_i=0$, and thus $F_i(0)=0$.

Similarly, to prove \Cref{eq:F_i_1}, consider the case when $p_i=1$. In this case, $T_i(j)=1$ if $j=i$ and $T_i(j)=0$ otherwise. Again, to ensure the unbiasedness of the reweighting function, we must have $\mathbb{E}_{E\sim P_E}[R_E(T_i)]=T_i$, which implies $R_E(T_i)=T_i$ almost surely. Therefore, $R_E(T_i)(i)=T_i(i)=p_i=1$, and thus $F_i(1)=1$.

To prove \Cref{eq:F_i_monotone}, consider two values $p_i\geq p'_i$. Define $T_i$ and $T'_i$ as follows:
\myvspace{-5pt}
\begin{equation*}
    T_i(j)=\begin{cases}
        p_i&j=i,\\
        1-p_i&j=i \opn{mod} n+1,\\
        0&\text{otherwise},
    \end{cases}
    \quad\quad
    T'_i(j)=\begin{cases}
        p'_i&j=i,\\
        1-p'_i&j=i \opn{mod} n+1,\\
        0&\text{otherwise}.
    \end{cases}
\end{equation*}
Since $p_i-p'_i=(T_i-T'_i)(i)\geq0$, by \Cref{eq:condition_2}, we have $F_i(p_i)-F_i(p'_i)=(R_E(T_i)-R_E(T'_i))(i)\geq0$, which proves the monotonicity of $F_i$.

To prove \Cref{eq:F_i_sum}, notice that due to \Cref{eq:R_E_P_i}, we have $\sum_i F_i(p_i) = \sum_i R_E(P)(i) = 1$.

Finally, according to \Cref{lem:function_equation}, the functions $F_i$ satisfying \Cref{eq:F_i_0,eq:F_i_1,eq:F_i_monotone,eq:F_i_sum} must be the identity function, i.e., $F_i(p)=p$ for all $i\in\{1,2,\dots,n\}$ and $p\in[0,1]$. Combining this with \Cref{eq:R_E_P_i}, we conclude that $R_E(P)=P$ almost surely for random $E$, which means that the reweighting function $R_E$ is trivial.

Therefore, when the vocabulary size $\abs{\Sigma}>2$, it is impossible to simultaneously maintain the watermark strength and sampling efficiency using non-trivial reweighting functions under the \textit{two reweight framework}.
\end{proof}  \myvspace{-10pt}
\section{Algorithms for Maintaining Watermark Strength or Sampling Efficiency}\label{se:maintain}
\myvspace{-7pt}

\begin{algorithm}\caption{Maintaining Watermark Strength
or Sampling Efficiency}\label{alg:both_algorithm}
\begin{algorithmic}
\STATE Given draft sequence length $K$, prompt $x_1,\dots,x_n$, target model $P(\cdot|\cdot)$, draft model $Q(\cdot|\cdot)$, code history $cch$ as a list of context code, context code function $cc:\Sigma^\ast\to C$, watermark code generation function $\hat{E}:C\times Z\to\mathcal{E}$, reweighting functions $R:\mathcal{E}\times\Delta_{\Sigma}\to\Delta_{\Sigma}$, and key for watermark $z\in Z$.
\STATE Initialize draft context code history $\widetilde{cch}\gets cch$.
\FOR{$t=1:K+1$}
\STATE Compute context code $c_t= cc(x_1,\dots,x_n,\tilde{x}_1,\dots,\tilde{x}_{t-1})$, watermark code $E_t=\hat{E}(c_t, z)$.
\STATE Check skipped $skipped_t=\begin{cases}
    \text{true}&c_t\text{~exists~in~}\widetilde{cch},\\
    \text{false}&c_t\text{~doesn't~exists~in~}\widetilde{cch}.
\end{cases}$
Set $\widetilde{cch}\gets \widetilde{cch}+[c_t]$.
\STATE \textbf{if} $t=K+1$ \textbf{then} Exit for loop. \textbf{end if}
\STATE Compute distribution $Q_t(\cdot)=Q(\cdot|x_1,\dots,x_n,\tilde{x}_1,\dots,\tilde{x}_{t-1})$.
\STATE Let $\mathfrak{Q}_t=\begin{cases}
    Q_t&skipped_t=\text{true},\\R_{E_t}(Q_t)&skipped_t=\text{false}.
\end{cases}$
Sample draft token $\tilde{x}_t\sim \mathfrak{Q}_t$.
\ENDFOR
\FOR{$t=1:K+1$ \textbf{in parallel}}
\STATE Compute distribution $P_t(\cdot)=P(\cdot|x_1,\dots,x_n,\tilde{x}_1,\dots,\tilde{x}_{t-1})$.
\STATE Let $\mathfrak{P}_t=\begin{cases}
    P_t&skipped_t=\text{true},\\R_{E_t}(P_t)&skipped_t=\text{false}.
\end{cases}$
\ENDFOR
\STATE Initialize empty output list: $out\gets []$.
Let $(\mathbb{P}_t,\mathbb{Q}_t)=\begin{cases}
    (\mathfrak{P}_t,\mathfrak{Q}_t)&\text{maintain watermark strength},\\(P_t,Q_t)&\text{maintain sampling efficiency}.
\end{cases}$
\FOR{$t=1:K$}
\STATE Set $cch\gets cch+[c_t]$.
Sample $r\sim U[0, 1]$ from a uniform distribution.
\STATE \textbf{if} $r<\min(1,\frac{\mathbb{P}_t(\tilde{x}_t)}{\mathbb{Q}_t(\tilde{x}_t)})$ \textbf{then} Set $out\gets out+[\tilde{x}_t]$. \textbf{else}
\begin{ALC@g}
\STATE Sample $x_{n+t}\sim(\mathbb{P}_t-\mathbb{Q}_t)_+$.
Set $out\gets out+[x_{n+t}]$.
Exit for loop.
\end{ALC@g}
\STATE \textbf{end if}
\ENDFOR
\IF{$out=[\tilde{x}_1,\dots,\tilde{x}_{K}]$}
\STATE Set $cch\gets cch+[c_{K+1}]$.
Sample $x_{n+K+1}\sim \mathfrak{P}_{K+1}$.
Set $out\gets out+[x_{n+K+1}]$.
\ENDIF
\STATE Return $out$ as generated tokens, and $cch$ as context code history.
\end{algorithmic}
\end{algorithm} 
In this section, we present two algorithms that aim to maintain either the watermark strength or the sampling efficiency under the \textit{two reweight framework}. In light of the no-go theorem (\Cref{thm:no_go}), which precludes the simultaneous maintenance of watermark strength and sampling efficiency, these algorithms provide deeper insights into the trade-offs between the two objectives.

\myvspace{-10pt}
\subsection{Maintaining Watermark Strength}
\myvspace{-7pt}

To maintain the watermark strength, we choose the reweight function for draft distribution to be the same as the reweight function  for the target distribution, i.e., $R'_E(Q)=R_E(Q)$. The speculative process is designed as follows:
\myvspace{-4pt}
\begin{equation}\label{eq:maintain_watermark_A}
    A(j|i)=\begin{cases}
        \min(1,\frac{R_E(P)(i)}{R_E(Q)(i)})&\text{if }i=j,\\
        \frac{(1-\frac{R_E(P)(i)}{R_E(Q)(i)})_+(R_E(P)(j)-R_E(Q)(j))_+}{\sum_{z\in\Sigma}(R_E(Q)(z)-R_E(P)(z))_+}&\text{if }i\ne j.
    \end{cases}
\end{equation}

\begin{theorem}[Maintaining Watermark Strength]\label{thm:maintain_watermark}
Under the \textit{two reweight framework}, if $R'_E(Q)=R_E(Q)$ and the speculative process $A(j|i)$ is defined as in \Cref{eq:maintain_watermark_A}, then the watermark strength is maintained, i.e., $\widehat{R}_E(P)=R_E(P)$. Moreover, the generation distribution is unbiased, i.e., $\mathbb{E}_{E\sim P_E}[\widehat{R}_E(P)]=P$ for all $P\in\Delta_\Sigma$.
\end{theorem} 
Intuitively, this algorithm applies the same reweighting function $R_E$ to both the draft distribution $Q$ and the target distribution $P$, and then performs speculative sampling based on the reweighted distributions $R_E(Q)$ and $R_E(P)$ as draft and target distribution.

\myvspace{-10pt}
\subsection{Maintaining Sampling Efficiency}
\myvspace{-10pt}

To maintain the sampling efficiency, we again choose the reweight function for draft distribution to be the same as the reweight function  for the target distribution, i.e., $R'_E(Q)=R_E(Q)$. However, the speculative process is designed differently:
\myvspace{-10pt}
\begin{equation}\label{eq:maintain_efficiency_A}
    A(j|i)=\begin{cases}
        \min(1,\frac{P(j)}{Q(i)})&\text{if }i=j,\\
        \frac{(1-\frac{P(i)}{Q(i)})_+(P(j)-Q(j))_+}{\sum_{z\in\Sigma}(Q(z)-P(z))_+}&\text{if }i\ne j.
    \end{cases}
\end{equation}

\begin{theorem}[Maintaining Sampling Efficiency]\label{thm:maintain_efficiency}
Under the \textit{two reweight framework}, if $R'_E(Q)=R_E(Q)$ and the speculative process $A(j|i)$ is defined as in \Cref{eq:maintain_efficiency_A}, then the sampling efficiency is maintained, i.e., $\alpha(P,Q)=\mathbb{E}_{E\sim P_E} [\sum_{i\in\Sigma}A(i|i)R'_E(Q)(i)]$. Moreover, the generation distribution is unbiased, i.e., $\mathbb{E}_{E\sim P_E}[\widehat{R}_E(P)]=P$ for all $P\in\Delta_\Sigma$.
\end{theorem} 
Intuitively, this algorithm generates a watermarked draft token using the watermarked draft distribution $R_E(Q)$, and then performs the standard speculative sampling process using the original distributions $Q$ and $P$ as draft and target distribution.

\subsection{Algorithms}

The pseudo code for the two methods described in the previous sections is provided in \Cref{alg:both_algorithm}. 
This pseudo code applies the methods in previous sections for multiple times in each step, and also considers the context code history to ensure unbiasedness for the whole sequence.
For reference, similar pseudo code for basic sampling, vanilla speculative sampling and vanilla unbiased watermarking is provided in \Cref{alg:basic,alg:sps,alg:wm}. 

\begin{remark}[Context code history]
According to \cite{hu2023unbiased}, the context code history is crucial for ensuring the unbiasedness of the entire generated sequence. In both algorithms, all accepted draft tokens' context codes need to be preserved in the context code history. Additionally, when a draft token is rejected, its context code should also be preserved because the newly generated random token after rejection, i.e. $x_{n+t}$, is not independent of the rejected random draft token $\tilde{x}_{t}$. By preserving the right context code history, we ensures that not only the distribution of a single token, but also the distribution of the entire generated sequence is unbiased.
During computing watermark score for detection, a context code history is also necessary so that each context code only contributes to the watermark score once.
\end{remark}
 \section{Experiments}\label{se:experiment}

To verify that \Cref{alg:both_algorithm} can indeed maintain either the watermark strength or the sampling efficiency as claimed, we test different methods on a text summarization task on \textsc{cnn\_dailymail} dataset \citep{see-etal-2017-get,DBLP:conf/nips/HermannKGEKSB15} using the Llama-7b model \citep{touvron2023llama} as the target model and the Llama-68m model \citep{miao2023specinfer} as the draft model.

We measure the sampling efficiency by the number of accepted tokens in the $out$ list of \Cref{alg:both_algorithm}, and report the Average Accepted Tokens Per Step (AATPS). A higher AATPS indicates a higher sampling efficiency. 

To measure the watermark strength, we compute the log P-value. For likelihood-based scores, the computation follows the method in \citep{hu2023unbiased}. For likelihood-agnostic scores, we use U score with the Chernoff bound in \Cref{eq:p_value}, where $\lambda$ is optimized numerically. We test the watermark strength for both the DeltaGumbel reweight and the Gamma reweight schemes. The Average Negative Log P-value Per Token (ANLPPT) is reported, with a higher value indicating a stronger watermark.

\begin{figure*}
\myvspace{-20pt}
\centering
\begin{subfigure}[b]{\textwidth}
  \centering
  \caption{DeltaGumbel Reweight}
  \begin{subfigure}[b]{0.45\textwidth}
  \includegraphics[height=5.3cm]{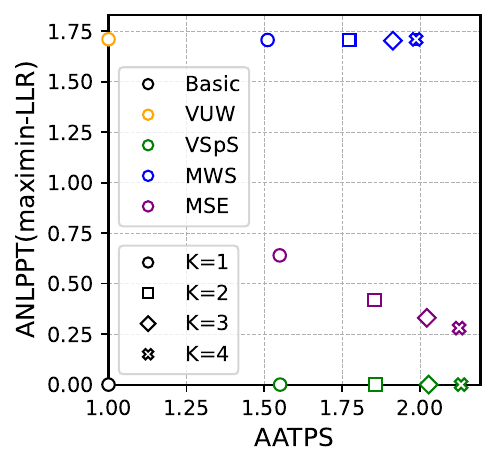}
  \end{subfigure}
  \begin{subfigure}[b]{0.45\textwidth}
  \includegraphics[height=5.3cm]{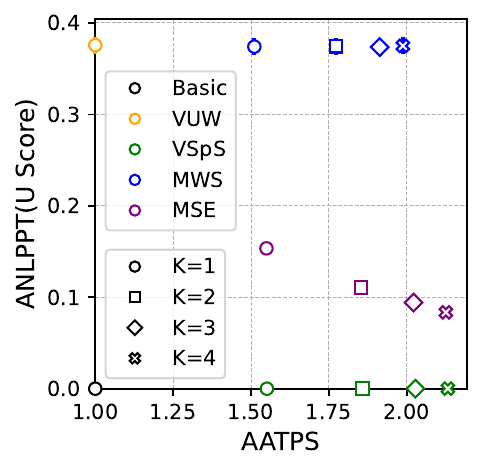}
  \end{subfigure}
\end{subfigure}
\begin{subfigure}[b]{\textwidth}
  \centering
  \caption{Gamma Reweight}
  \begin{subfigure}[b]{0.45\textwidth}
  \includegraphics[height=5.3cm]{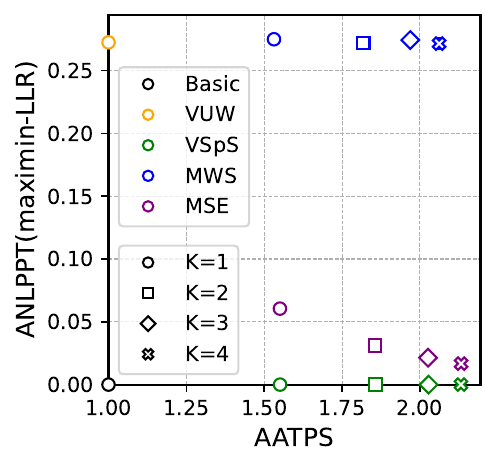}
  \end{subfigure}
  \begin{subfigure}[b]{0.45\textwidth}
  \includegraphics[height=5.3cm]{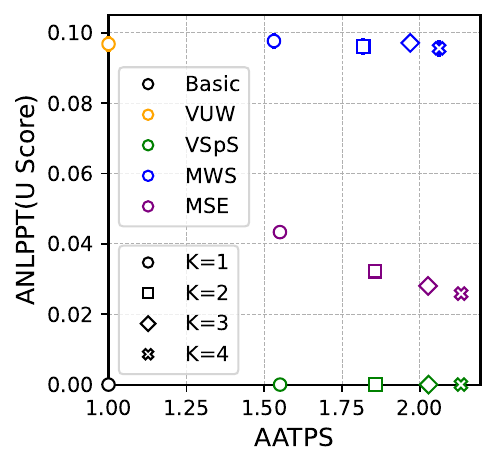}
  \end{subfigure}
\end{subfigure}
\caption{
Comparison of different methods.
The x-axis shows the Average Accepted Tokens Per Step (AATPS) as a measure of speculative sampling efficiency, while y-axis shows the Average Negative Log P-value Per Token (ANLPPT) as a measure of watermark strength. The P-value is computed based on either a likelihood-based test using the maximin-LLR score (left) or a likelihood-agnostic test using the U score (right). Watermarking is performed using either the DeltaGumbel reweight (top) or the Gamma reweight (bottom). Error bars represent $3\sigma$ confidence intervals\protect\footnotemark.
}\label{fig:verify}
\myvspace{-10pt}
\end{figure*}

\footnotetext{The error bars for some methods are very small and may not be visible in the plot. The exact error bar can be found in \Cref{tab:table1_summarization_scan_n_llama-7b_llama-68m}.} 
The results are shown in \Cref{fig:verify}. We compare the performance of Basic Sampling, Vanilla Unbiased Watermark (VUW), Vanilla Speculative Sampling (VSpS), Maintain Watermark Strength (MWS), and Maintain Sampling Efficiency (MSE).

We also measure the Per Token Time (PTT) in millisecond to evaluate the wall-time latency and verify that \Cref{alg:both_algorithm} can indeed achieve acceleration compared to the vanilla unbiased watermark method. The Log Perplexity (LOGPPL) is computed to verify that all algorithms produce the same output distribution and do not affect the quality of the language model output. The raw data for these additional metrics are provided in \Cref{tab:table1_summarization_scan_n_llama-7b_llama-68m} in the appendix due to space constraints.

We also conduct additional experiments using different models and tasks. In addition to the Llama-7b model, we test the Llama-13b model \citep{touvron2023llama} as the target model, with Llama-68m \citep{miao2023specinfer} as the draft model. Besides the text summarization task, we also evaluate the methods on an open-ended text generation task. The results of these additional experiments are provided in \Cref{se:addtionalexperiment}.
The total computational cost for reproducing all the experiments in this paper is approximately 1200 A6000 GPU hours.

The experimental results in \Cref{fig:verify} and \Cref{se:addtionalexperiment} support the following findings:

\begin{itemize}[leftmargin=0.15in]
\item \Cref{alg:both_algorithm} can indeed maintain either the watermark strength or the sampling efficiency as claimed. The MWS method achieves the same watermark strength as the VUW method, while the MSE method achieves the same sampling efficiency as the VSpS method.
\item \Cref{alg:both_algorithm} can indeed accelerate the generation process compared to the vanilla unbiased watermark method. Both the MWS and MSE methods achieve lower PTT than the VUW method, as shown in \Cref{tab:table1_summarization_scan_n_llama-7b_llama-68m}. 
\item MWS method has only marginal sampling efficiency gap compared to VSpS, while maintain the watermark strength as VUW method, making it highly practical.
\item All algorithms produce the same output distribution and do not affect the quality of the language model output, as evidenced by the similar LOGPPL values across all methods in \Cref{tab:table1_summarization_scan_n_llama-7b_llama-68m}.
\item The above findings are consistent across different draft sequence length ($K=1,2,3,4$), different models (Llama-7b and Llama-13b), different tasks (text summarization and open-ended text generation), different reweight schemes (DeltaGumbel and Gamma), and different watermark detection methods (likelihood-based and likelihood-agnostic). Our extensive experiments validate the generality of the findings.
\end{itemize}

In summary, our experimental results validate the theoretical findings and demonstrate the effectiveness of the proposed \Cref{alg:both_algorithm}.

 \myvspace{-2pt}
\section{Conclusion}
\myvspace{-3pt}
Our work provides a rigorous theoretical foundation for understanding the trade-off between watermark strength and sampling efficiency in the context of accelerated generation of watermarked tokens from large language models. We prove a no-go theorem, showing that non-trivial trade-offs are inevitable when the vocabulary size is greater than two. To explore these trade-offs, we 
design algorithms that prioritize either watermark strength or sampling efficiency. Our findings contribute to the development of methods for protecting the intellectual property of language models while leveraging the efficiency of speculative sampling techniques. 
\section*{Acknowledgments}

ZH and HH were partially supported by NSF IIS 2347592, 2347604, 2348159, 2348169, DBI 2405416, CCF 2348306, CNS 2347617. 
\bibliography{content/references}
\bibliographystyle{plainnat}

\newpage
\appendix
\onecolumn
\section{Related Works}\label{se:related}
Our work lies at the intersection of two active research areas: speculative sampling for accelerated inference and watermarking techniques for language models.

\subsection{Speculative Sampling for Accelerated Inference}
In the domain of speculative sampling, a common approach is to use a smaller language model as the draft model \citep{leviathan2023fast,chen2023accelerating}. Efforts have been made to further increase the overlap between the draft and target models through distillation \citep{zhou2023distillspec}. Other works focus on modifying the target model itself, such as adding ``look ahead'' tokens \citep{monea2023pass}, introducing new heads to predict future tokens \citep{cai2024medusa}, reusing the computation of the large model to achieve a better latency-overlap trade-off for the draft model \citep{li2024eagle}, or using the target model with partial key-value cache as the draft model \citep{sun2024triforce}. Alternative approaches include using document retrieval \citep{yang2023inference,he2023rest} or n-gram models \citep{ou2024lossless} as the draft model.

When the draft sequence length is greater than one, vanilla speculative sampling is known to be suboptimal. Methods have been proposed to amend the verification process for the draft sequence, verifying the entire sequence at once instead of individual tokens, leading to a longer expected number of accepted tokens \citep{sun2024optimal}.

An extension of speculative sampling is to change the sequence input to a tree input. While typical language models take a sequence as input, which is a path in the symbol tree space, some works modify the input to be a tree with multiple branches. A single forward pass can then obtain probabilities on multiple branches, gathering more information to help accelerate decoding. This requires modifying the transformer implementation to change the causal attention to tree attention \citep{yang2024multi,miao2023specinfer,cai2024medusa,spector2023accelerating}. Speculative sampling can also be used repeatedly, with an additional draft model to accelerate the draft model itself \citep{spector2023accelerating}.

Our work is independent of the specific draft model used. While many recent advancements stem from faster and more accurate draft models, our method does not rely on any assumptions about the draft model. A better draft model can always be plugged in to provide faster acceleration. 

The methods in \Cref{se:maintain} of the main text only consider the basic speculative sampling approach and do not take into account other variants such as verifying the entire sequence, tree verification, or multi-candidacy. However, our ideas can be extended to these variants and still maintain either the watermark strength or the sampling efficiency, as discussed in \Cref{se:extension}.

\subsection{Watermarking Techniques for Language Models}

In the domain of watermarking for language models, various approaches have been explored. Some works attempt to edit existing text to embed watermarks \citep{brassil1995electronic,por2012unispach,rizzo2016content,sato2023embarrassingly,topkara2006hiding,munyer2023deeptextmark,yang2022tracing,yang2023watermarking,yoo2023robust,atallah2001natural,topkara2006words,meral2009natural}. Others try to incorporate watermarks during the training phase \citep{liu2023watermarking,tang2023did,sun2022coprotector,sun2023codemark}.

More closely related to our work is the direction of modifying the sampling stage to directly generate watermarked results. Since the pioneering works of \citet{aaron2022blog} and \citet{kirchenbauer2023watermark}, watermarking techniques have seen significant development. 

To address the bias introduced by watermarking, researchers have proposed skipping watermarking on low-entropy tokens \citep{liu2024adaptive,wang2023towards} or accumulating entropy during the generation process and only adding a watermark when the accumulated entropy exceeds a threshold \cite{christ2023undetectable}. \citet{hu2023unbiased} introduced a framework that includes unbiased reweighting and context code history to ensure that the output distribution is strictly unbiased.

Subsequently, many variants have been proposed, including multi-bit watermarks \citep{yoo2023advancing,fernandez2023three} and more robust watermarking schemes \citep{ren2023robust,liu2023semantic,kirchenbauer2023reliability,zhao2023provable,kuditipudi2023robust,hou2023semstamp}. Efforts have also been made to search for better watermark detection methods \citep{wu2023dipmark,li2024statistical}.

Our work builds upon the unbiased watermarking framework of \citet{hu2023unbiased} and explores the trade-off between watermark strength and sampling efficiency when integrating watermarking with speculative sampling. To the best of our knowledge, this is the first work to investigate this intersection and provide theoretical insights and practical algorithms for navigating the inherent trade-offs.
 \section{Proofs}\label{se:proofs}
\begin{proof}[Proof of \Cref{lem:unbiased_draft}]
Let $P=Q$. Then we have
\begin{equation*}
    1=\alpha(P,Q)=\mathbb{E}_{E\sim P_E} \left[\sum_{i\in\Sigma}A(i|i)R'_E(Q)(i)\right].
\end{equation*}
Note that for all $i\in\Sigma$, $A(i|i)\leq1$, and thus
\begin{equation*}
    \sum_{i\in\Sigma}A(i|i)R'_E(Q)(i)\leq\sum_{i\in\Sigma}R'_E(Q)(i)=1.
\end{equation*}
Therefore, we must have $A(i|i)=1$ almost surely for random $E$ and for all $i\in\Sigma$.
Considering the unbiasedness requirement for the final output distribution, i.e.,
\begin{equation}\label{eq:unbiased_output}
    \forall P\in\Delta_\Sigma,\mathbb{E}_{E\sim P_E}[A\circ R'_E(Q)]=P,
\end{equation}
we obtain $\mathbb{E}_{E\sim P_E}[R'_E(Q)]=P=Q$.
\end{proof} \begin{proof}[Proof of \Cref{lem:same_reweight}]
Let $P=Q$. Following the proof of \Cref{lem:unbiased_draft}, we have $A(i|i)=1$ almost surely for random $E$ and for all $i\in\Sigma$. Therefore,
\begin{equation*}
    \widehat{R}_E(P)=A\circ R'_E(Q)=R'_E(Q).
\end{equation*}
Since $\widehat{R}_E(P)=R_E(P)$, we conclude that $R'_E(Q)=R_E(Q)$.
\end{proof} 
\begin{lemma}[A Function Equation]\label{lem:function_equation}
Given $n$ monotonically increasing functions $F_i:[0,1]\to[0,1]$ for $i\in\{1,2,3,\dots,n\}$, i.e., $x\geq x'\implies F_i(x)\geq F_i(x')$, satisfying
\begin{gather*}
    \forall i\in\{1,2,3,\dots,n\},F_i(0)=0,F_i(1)=1,\\
    \sum_i x_i=1\implies \sum_i F_i(x_i)=1,
\end{gather*}
we have $F_1(x)=F_i(x)=x$ for all $i\in\{1,2,3,\dots,n\}$ and $x\in[0,1]$.
\end{lemma} \begin{proof}[Proof of \Cref{lem:function_equation}]
We first prove that $F_1(x)=F_2(x)$ for all $x\in[0,1]$. Let $x_1=0$, $x_2=1-x_3$, and $x_i=0$ for all $i\geq4$. We obtain
\begin{equation*}
    F_3(x_3)=1-F_2(1-x_3).
\end{equation*}
Next, let $x_2=0$, $x_3=1-x_1$, and $x_i=0$ for all $i\geq4$. This gives us
\begin{align*}
    F_1(x_1)&=1-F_3(1-x_1)\\
            &=1-(1-F_2(1-(1-x_1)))\\
            &=F_2(x_1).
\end{align*}
Similarly, we can show that $F_1(x)=F_i(x)$ for all $i\in\{1,2,3,\dots,n\}$ and $x\in[0,1]$.

Next, we prove that for all $n\in\mathbb{N}$ and $b\leq 2^n$, $F_1(\frac{b}{2^n})=\frac{b}{2^n}$. First, let $x_1=\frac{1}{2}$, $x_2=\frac{1}{2}$, and $x_i=0$ for all $i\geq3$. We obtain $F_1(\frac{1}{2})=\frac{1}{2}$. 

Assume that for some $n$, we have $F_1(\frac{b}{2^n})=\frac{b}{2^n}$ for all $b\leq 2^n$. We will prove that $F_1(\frac{b}{2^{n+1}})=\frac{b}{2^{n+1}}$ for all $b\leq 2^{n+1}$. 

For $b\leq 2^n$, let $x_1=\frac{b}{2^{n+1}}$, $x_2=\frac{b}{2^{n+1}}$, $x_3=1-\frac{b}{2^{n}}$, and $x_i=0$ for all $i\geq4$. We obtain $F_1(\frac{b}{2^{n+1}})=\frac{b}{2^{n+1}}$. 

For $2^n\leq b\leq 2^{n+1}$, let $x_1=\frac{b}{2^{n+1}}$, $x_2=1-\frac{b}{2^{n+1}}$, and $x_i=0$ for all $i\geq3$. We obtain $F_1(\frac{b}{2^{n+1}})=\frac{b}{2^{n+1}}$. 

By mathematical induction, we have $F_1(\frac{b}{2^n})=\frac{b}{2^n}$ for all $n\in\mathbb{N}$ and $b\leq 2^n$.

Since $F_1$ is monotonically increasing, for all $x\in[0,1]$ and $n\in\mathbb{N}$, we have
\begin{equation*}
    F_1(\frac{\lfloor{x2^n}\rfloor}{2^n})\leq F_1(x)\leq F_1(\frac{\lceil{x2^n}\rceil}{2^n}).
\end{equation*}
Taking the limit as $n\to\infty$, we obtain
\begin{align*}
    \lim_{n\to\infty} F_1(\frac{\lfloor{x2^n}\rfloor}{2^n})&=x,\\
    \lim_{n\to\infty} F_1(\frac{\lceil{x2^n}\rceil}{2^n})&=x.
\end{align*}
Therefore, $F_1(x)=x$ for all $x\in[0,1]$, and consequently, $F_i(x)=x$ for all $i\in\{1,2,3,\dots,n\}$ and $x\in[0,1]$.
\end{proof} 
\begin{proof}[Proof of \Cref{thm:maintain_watermark}]
First, we have $\widehat{R}_E(P)=(A\circ R'_E)(Q)=(A\circ R_E)(Q)$. For any $j\in\Sigma$,
\begin{align*}
    \widehat{R}_E(P)(j)
    =&\sum_i A(j|i) R_E(Q)(i)\\
    =&\min(R_E(Q)(j),R_E(P)(j))\\&+\sum_{i\neq j} \frac{(R_E(Q)(i)-R_E(P)(i))_+(R_E(P)(j)-R_E(Q)(j))_+}{\sum_{z\in\Sigma}(R_E(Q)(z)-R_E(P)(z))_+}\\
    =&\min(R_E(Q)(j),R_E(P)(j))+(R_E(P)(j)-R_E(Q)(j))_+\\
    =&R_E(P)(j).
\end{align*}
Therefore, $\widehat{R}_E(P)=R_E(P)$, which means the watermark strength is maintained.

To prove the unbiasedness of the generation distribution, note that for all $P\in\Delta_\Sigma$,
\begin{equation*}
    \mathbb{E}_{E\sim P_E}[\widehat{R}_E(P)]=\mathbb{E}_{E\sim P_E}[R_E(P)]=P,
\end{equation*}
where the last equality follows from the unbiasedness of the reweighting function $R_E$.
\end{proof} \begin{proof}[Proof of \Cref{thm:maintain_efficiency}]
To prove that the sampling efficiency is maintained, we have
\begin{align*}
    \mathbb{E}_{E\sim P_E} \left[\sum_{i\in\Sigma}A(i|i)R'_E(Q)(i)\right]
    &=\mathbb{E}_{E\sim P_E} \left[\sum_{i\in\Sigma}A(i|i)R_E(Q)(i)\right]\\
    &=\sum_{i\in\Sigma}A(i|i)\mathbb{E}_{E\sim P_E} [R_E(Q)(i)]\\
    &=\sum_{i\in\Sigma}A(i|i)Q(i)\\
    &=\sum_{i\in\Sigma}\min(Q(i),P(i))\\
    &=\alpha(P,Q).
\end{align*}
To prove the unbiasedness of the generation distribution, we have for any $j\in\Sigma$,
\begin{align*}
    \mathbb{E}_{E\sim P_E}[\widehat{R}_E(P)](j)
    &=\mathbb{E}_{E\sim P_E}\left[\sum_i A(j|i)R_E(Q)(i)\right]\\
    &=\sum_i A(j|i)\mathbb{E}_{E\sim P_E}[R_E(Q)](i)\\
    &=\sum_i A(j|i)Q(i)\\
    &=\min(Q(j),P(j))+\sum_{i\neq j}\frac{(Q(i)-P(i))_+(P(j)-Q(j))_+}{\sum_{z\in\Sigma}(Q(z)-P(z))_+}\\
    &=\min(Q(j),P(j))+(P(j)-Q(j))_+\\
    &=P(j).
\end{align*}
Therefore, $\mathbb{E}_{E\sim P_E}[\widehat{R}_E(P)]=P$ for all $P\in\Delta_\Sigma$, which means the generation distribution is unbiased.
\end{proof}  \section{Algorithms}\label{se:algorithms}
\begin{algorithm}[H]
\caption{Basic Sampling}\label{alg:basic}
\begin{algorithmic}
\STATE Given generated sequence length $K$, prompt $x_1,\dots,x_n$, target model $P(\cdot|\cdot)$.
\STATE Initialize empty output list: $out\gets []$.
\FOR{$t=1:K$}
\STATE Compute distribution $P_t(\cdot)=P(\cdot|x_1,\dots,x_n,x_{n+1},\dots,x_{n+t-1})$.
\STATE Sample token $x_{n+t}\sim P_t$.
Set $out\gets out+[x_{n+t}]$.
\ENDFOR
\STATE Return $out$ as generated tokens.
\end{algorithmic}
\end{algorithm} \begin{algorithm}[H]
\caption{Vanilla Speculative Sampling}\label{alg:sps}
\begin{algorithmic}
\STATE Given draft sequence length $K$, prompt $x_1,\dots,x_n$, target model $P(\cdot|\cdot)$, and draft model $Q(\cdot|\cdot)$.
\FOR{$t=1:K$}
\STATE Compute distribution $Q_t(\cdot)=Q(\cdot|x_1,\dots,x_n,\tilde{x}_1,\dots,\tilde{x}_{t-1})$.
Sample draft token $\tilde{x}_t\sim Q_t$.
\ENDFOR
\FOR{$t=1:K+1$ \textbf{in parallel}}
\STATE Compute distribution $P_t(\cdot)=P(\cdot|x_1,\dots,x_n,\tilde{x}_1,\dots,\tilde{x}_{t-1})$.
\ENDFOR
\STATE Initialize empty output list: $out\gets []$.
\FOR{$t=1:K$}
\STATE Sample $r\sim U[0, 1]$ from a uniform distribution.
\STATE \textbf{if} $r<\min(1,\frac{P_t(\tilde{x}_t)}{Q_t(\tilde{x}_t)})$ \textbf{then} Set $out\gets out+[\tilde{x}_t]$. \textbf{else}
\begin{ALC@g}
\STATE Sample $x_{n+t}\sim(P_t-Q_t)_+$.
Set $out\gets out+[x_{n+t}]$.
Exit for loop.
\end{ALC@g}
\STATE \textbf{end if}
\ENDFOR
\IF{$out=[\tilde{x}_1,\dots,\tilde{x}_{K}]$}
\STATE Sample $x_{n+K+1}\sim P_{K+1}$.
Set $out\gets out+[x_{n+K+1}]$.
\ENDIF
\STATE Return $out$ as generated tokens.
\end{algorithmic}
\end{algorithm} \begin{algorithm}[H]
\caption{Vanilla Unbiased Watermark Method}\label{alg:wm}
\begin{algorithmic}
\STATE Given generated sequence length $K$, prompt $x_1,\dots,x_n$, target model $P(\cdot|\cdot)$, code history $cch$ as a list of context code, context code function $cc:\Sigma^\ast\to C$, watermark code generation function $\hat{E}:C\times Z\to\mathcal{E}$, reweighting functions $R:\mathcal{E}\times\Delta_{\Sigma}\to\Delta_{\Sigma}$, and key for watermark $z\in Z$.
\STATE Initialize empty output list: $out\gets []$.
\FOR{$t=1:K$}
\STATE Compute context code $c_t= cc(x_1,\dots,x_n,x_{n+1},\dots,x_{n+t-1})$, watermark code $E_t=\hat{E}(c_t, z)$.
\STATE Check skipped $skipped_t=\begin{cases}
    \text{true}&c_t\text{~exists~in~}cch,\\
    \text{false}&c_t\text{~doesn't~exists~in~}cch.
\end{cases}$
Set $cch\gets cch+[c_t]$.
\STATE Compute distribution $P_t(\cdot)=P(\cdot|x_1,\dots,x_n,x_{n+1},\dots,x_{n+t-1})$.
\STATE Let $\mathfrak{P}_t=\begin{cases}
    P_t&skipped_t=\text{true},\\R_{E_t}(P_t)&skipped_t=\text{false}.
\end{cases}$
Sample token $x_{n+t}\sim \mathfrak{P}_t$.
Set $out\gets out+[x_{n+t}]$.
\ENDFOR
\STATE Return $out$ as generated tokens.
\end{algorithmic}
\end{algorithm}  \section{U Score, DeltaGubel Reweight, and Gamma Reweight}\label{se:Uscore}
In this section, we provide detailed definitions of the U score, DeltaGumbel reweight, and Gamma reweight schemes, which are used in the main text to compute the P-value for detecting watermarks.

\subsection{DeltaGumbel Reweight}
In the DeltaGumbel reweight scheme, the watermark code $E$ is a list of $\abs{\Sigma}$ independent and identically distributed standard Gumbel variables. The reweighting function is defined as:
\begin{equation}
    R_E(P):=\delta_{a^\ast}, \quad\quad
    a^*:=\operatorname{argmax}_a\{\log P(a)\:+\:E(a)\}
\end{equation}
where $\delta_{a^\ast}$ is the Dirac delta function centered at $a^\ast$.

The U score for the DeltaGumbel reweight is defined as:
\begin{equation}
    U=\exp(-\exp(-E(x)))\in[0,1].
\end{equation}

If there is no watermark added while generating $x$, in other word, if the token $x$ is independent with $E$, then the random $U$ is uniformly distribution in $[0,1]$.

The logarithm of the moment-generating function (MGF) of the U score for the DeltaGumbel reweight is given by:
\begin{equation}
    \log\mathbb{E}[\exp(\lambda U)]=- \log{\left(\lambda \right)} + \log{\left(e^{\lambda} - 1 \right)}.
\end{equation}

\subsection{Gamma Reweight}
In the Gamma reweight scheme, the watermark code $E$ is a random bijection from $\Sigma$ to the set $\{0,1,2,\dots,\abs{\Sigma}-1\}$. The reweighting function is defined as:
\begin{gather}
    R_E(P)(t):=A_{E,P}(E(t))-A_{E,P}(E(t)-1),\\
    A_{E,P}(i):=\max\left\{2\left(\sum_{a\in\Sigma}\mathbf{1}(E(a)\leq i)P(a)\right)-1,0\right\}.
\end{gather}

The U score for the Gamma reweight is defined as:
\begin{equation}
    U=\frac{E(x)+\frac{1}{2}}{\abs{\Sigma}}\in[0,1].
\end{equation}

If there is no watermark added while generating $x$, in other word, if the token $x$ is independent with $E$, then the random $U$ is uniformly distribution in $\{\frac{\frac{1}{2}}{\abs{\Sigma}},\frac{\frac{3}{2}}{\abs{\Sigma}},\dots,\frac{\abs{\Sigma}-\frac{1}{2}}{\abs{\Sigma}}\}$.

The logarithm of the moment-generating function (MGF) of the U score for the Gamma reweight is given by:
\begin{equation}
    \log\mathbb{E}[\exp(\lambda U)]=- \log{\left(2\abs{\Sigma}\opn{sinh}(\frac{\lambda}{2\abs{\Sigma}}) \right)} + \log{\left(e^{\lambda} - 1 \right)}.
\end{equation}

Both the DeltaGumbel reweight and Gamma reweight schemes are unbiased \citep{hu2023unbiased}, meaning that for any distribution $P\in\Delta_\Sigma$, we have:
\begin{equation}
    \mathbb{E}_{E\sim P_E}[R_E(P)]=P.
\end{equation}

The U scores defined for these reweight schemes are likelihood-agnostic, which means that they do not depend on the original distribution $P$. This property makes them possibly more robust to perturbations compared to likelihood-based scores such as the LLR score.

To compute the P-value for detecting watermarks using the U score, we can substitute the corresponding MGF into \Cref{eq:p_value}. \section{Extension to Variants of Speculative Sampling}\label{se:extension}

The analysis and process presented in \Cref{se:maintain} focus on the basic speculative sampling approach, where a single draft token is sampled and then accepted or rejected. 
\Cref{alg:both_algorithm} apply such process multiple times, accepting or rejecting tokens one by one, similar to vanilla speculative sampling.

Recent developments in speculative sampling have introduced various new techniques, such as verifying the entire sequence, tree verification, or multi-candidacy (see \Cref{se:related} for details). While \Cref{alg:both_algorithm} in \Cref{se:maintain} does not explicitly consider these variants, the underlying ideas can be directly extended to general speculative sampling approaches.

To maintain the watermark strength, the intuition is to apply the watermark to the draft tokens by sampling them from the watermarked draft model distribution, denoted as $Q_w$. \textbf{Then, the watermarked target distribution $P_w$ is computed, and speculative sampling is performed, treating $Q_w$ as the draft model and $P_w$ as the target model.} Since speculative sampling ensures that the generated content follows the distribution of the target model, the final generated results will be drawn from the distribution $P_w$. Consequently, the watermark strength remains unchanged compared to directly sampling from $P_w$.

To maintain the sampling efficiency, the intuition is to apply the watermark to the draft tokens by sampling them from the watermarked draft model distribution $Q_w$. \textbf{Then, speculative sampling is performed, treating $Q$ as the draft model and $P$ as the target model.} Under the expectation of random watermark codes, the draft tokens follow the distribution $Q$. Therefore, the efficiency of speculative sampling remains unchanged compared to directly sampling draft tokens from $Q$.

These arguments do not depend on the specific form of speculative sampling, and do not assume the structure of the draft tokens and draft models, making the ideas presented in \Cref{se:maintain} applicable to various speculative sampling variants.

We acknowledge that experimental validations would be helpful to demonstrate the effectiveness of the extended methods in practice. However, due to implementation/computation cost and the focus of this paper on the foundational theory, we only present a high-level discussion and left out the empirical validations for extended methods.

 \section{Broader Impacts}\label{se:impact}
This paper presents work whose goal is to accelerate the generation speed of existing watermarking methods for large language models. There are several potential positive societal impacts of our work. By making watermarking techniques more practical and efficient, it may encourage their wider adoption. This can help protect the intellectual property rights. 

However, there are also potential negative societal impacts to consider. Although our unbiased watermarking approach ensures the validity of the model outputs is not compromised, there is a risk that watermarking techniques could be abused. For example, unbiased watermarks are undetectable, which could enable tracking and surveillance, raising privacy concerns.

To mitigate potential negative impacts, it is important that watermarking techniques are used responsibly. This includes transparency about the use of watermarks, obtaining user consent where applicable, and putting safeguards in place to prevent misuse.

In conclusion, while our work on accelerating watermarking for language models has the potential to encourage wider adoption and protect intellectual property, it is important to carefully consider and address potential negative societal impacts to ensure the technology is used responsibly and ethically. \section{Limitation}\label{se:limitation}
Our work makes significant contributions to the field of watermarking and speculative sampling for large language models, but it also has several limitations.

In terms of theoretical analysis, the no-go theorem assumes a specific \textit{two reweight framework}. Although this framework is general, it is possible that other frameworks or methods may lead to different theoretical results. This paper represents the first exploration in this direction, and the \textit{two reweight framework} is also the first attempt. Future work may discover more powerful frameworks that yield different insights.

Regarding experimental validation, we use relatively small language models and basic draft model. While our experiments verify the effectiveness of the theory, the acceleration ratio may not represent the state-of-the-art. Speculative sampling techniques have been developing rapidly in recent times. If combined with the latest advances, it should be possible to achieve even higher Average Accepted Tokens Per Step (AATPS) and lower Per Token Time (PTT), though it is not directly related to our main contribution.

The choice of draft sequence length is critical in the deployment. Most of the existing methods select the optimal draft sequence length based on trial and error. This paper does not make contributions to determining the optimal draft sequence length. However, it should be noted that the maintain watermark strength (MWS) method proposed in this paper reduces the speculative efficiency, which also affects the selection of the optimal draft sequence length. Care should be taken in the deployment to optimally select this parameter.

If the maintain watermark strength (MWS) method is used, the inference speed will decrease slightly. This creates a side channel, and users may infer from this whether the backend service uses the MWS algorithm, which compromises the undetectability of the watermark.

Despite these limitations, our work provides valuable insights and advances the state-of-the-art in watermarking and speculative sampling techniques for large language models. We hope our findings will stimulate further research to realize the full potential of these techniques.
 \section{Additional Experiment Results}\label{se:addtionalexperiment}
\newpage

\begin{figure}[H]
\centering
\begin{subfigure}[b]{\textwidth}
  \centering
  \caption{DeltaGumbel Reweight}
  \begin{subfigure}[b]{0.45\textwidth}
  \includegraphics[height=5.3cm]{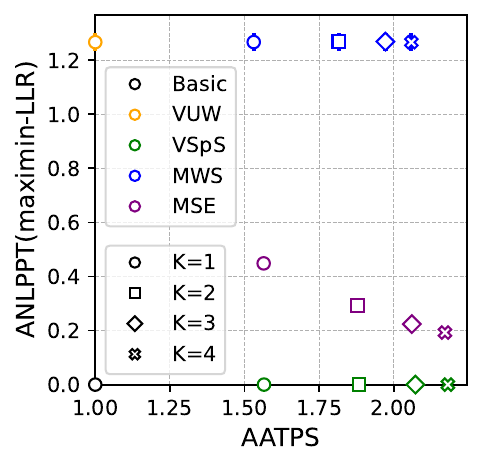}
  \end{subfigure}
  \begin{subfigure}[b]{0.45\textwidth}
  \includegraphics[height=5.3cm]{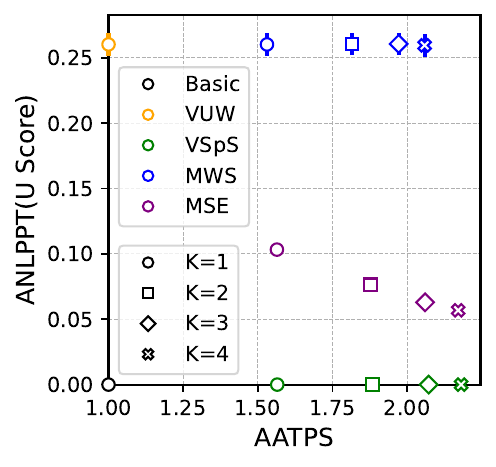}
  \end{subfigure}
\end{subfigure}
\begin{subfigure}[b]{\textwidth}
  \centering
  \caption{Gamma Reweight}
  \begin{subfigure}[b]{0.45\textwidth}
  \includegraphics[height=5.3cm]{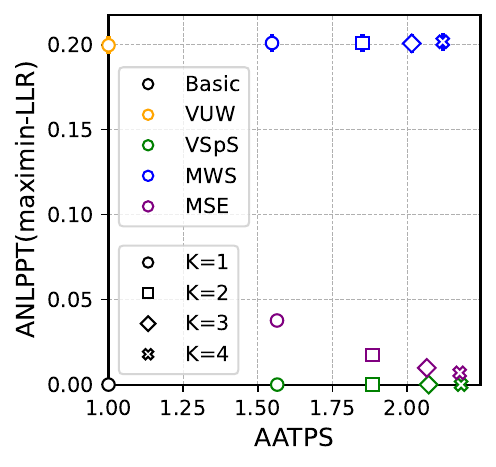}
  \end{subfigure}
  \begin{subfigure}[b]{0.45\textwidth}
  \includegraphics[height=5.3cm]{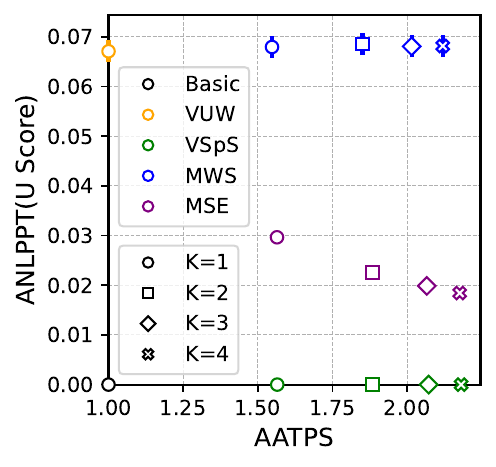}
  \end{subfigure}
\end{subfigure}
\caption{
Text summarization task with LLaMa-13b model \cite{touvron2023llama} as target model and LLaMa-68m model \cite{miao2023specinfer} as reference model.
}

\end{figure} \begin{figure}[H]
\centering
\begin{subfigure}[b]{\textwidth}
  \centering
  \caption{DeltaGumbel Reweight}
  \begin{subfigure}[b]{0.45\textwidth}
  \includegraphics[height=5.3cm]{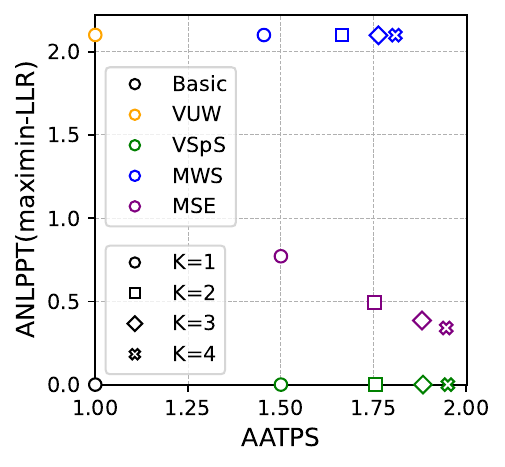}
  \end{subfigure}
  \begin{subfigure}[b]{0.45\textwidth}
  \includegraphics[height=5.3cm]{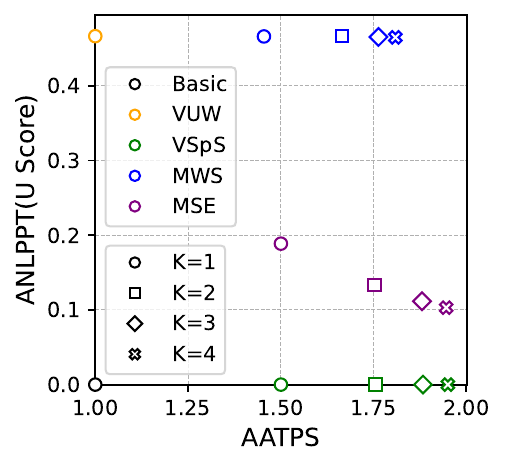}
  \end{subfigure}
\end{subfigure}
\begin{subfigure}[b]{\textwidth}
  \centering
  \caption{Gamma Reweight}
  \begin{subfigure}[b]{0.45\textwidth}
  \includegraphics[height=5.3cm]{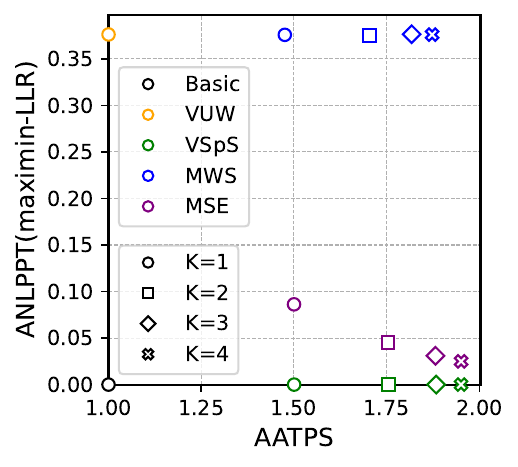}
  \end{subfigure}
  \begin{subfigure}[b]{0.45\textwidth}
  \includegraphics[height=5.3cm]{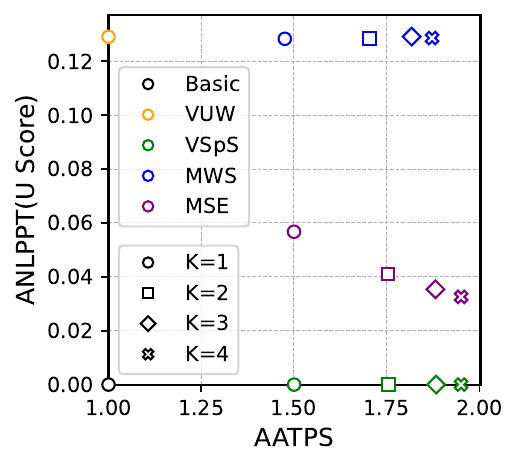}
  \end{subfigure}
\end{subfigure}
\caption{
Open-ended text generation task with LLaMa-7b model \cite{touvron2023llama} as target model and LLaMa-68m model \cite{miao2023specinfer} as reference model.
}

\end{figure} \begin{figure}[H]
\centering
\begin{subfigure}[b]{\textwidth}
  \centering
  \caption{DeltaGumbel Reweight}
  \begin{subfigure}[b]{0.45\textwidth}
  \includegraphics[height=5.3cm]{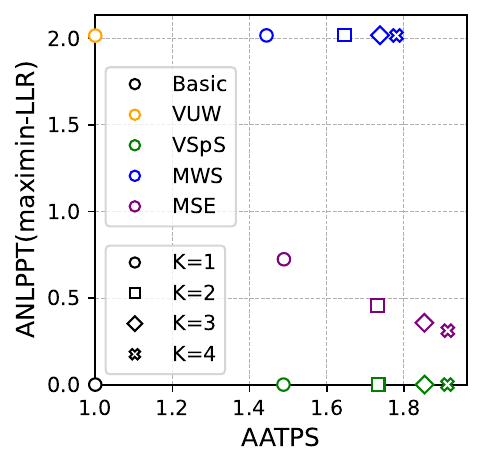}
  \end{subfigure}
  \begin{subfigure}[b]{0.45\textwidth}
  \includegraphics[height=5.3cm]{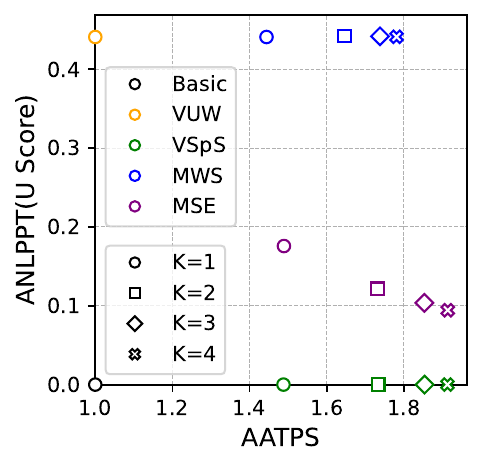}
  \end{subfigure}
\end{subfigure}
\begin{subfigure}[b]{\textwidth}
  \centering
  \caption{Gamma Reweight}
  \begin{subfigure}[b]{0.45\textwidth}
  \includegraphics[height=5.3cm]{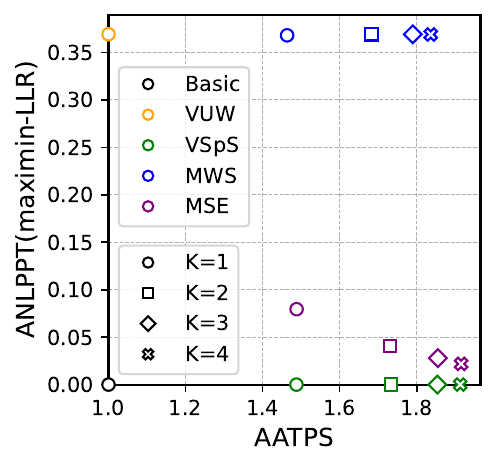}
  \end{subfigure}
  \begin{subfigure}[b]{0.45\textwidth}
  \includegraphics[height=5.3cm]{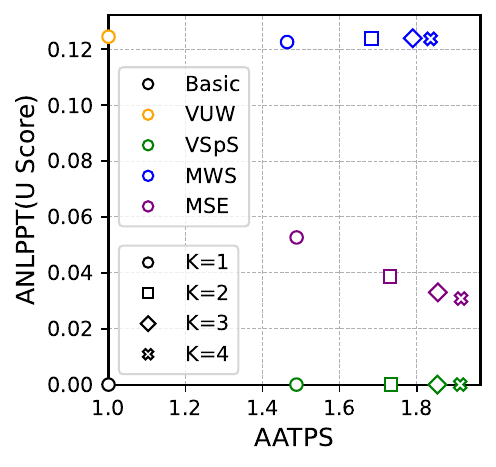}
  \end{subfigure}
\end{subfigure}
\caption{
Open-ended text generation task with LLaMa-13b model \cite{touvron2023llama} as target model and LLaMa-68m model \cite{miao2023specinfer} as reference model.
}

\end{figure} 
\begin{table}[H]
\centering
\begin{tabular}{rlllll}
\toprule
$K$ & method & reweight & AATPS & PTT & LOGPPL \\
\midrule
\midrule
1 & Basic & No Reweight & $1.0\pm0.0$ & $29.58\pm0.07$ & $1.75\pm0.03$ \\
1 & VUW & DeltaGumbel & $1.0\pm0.0$ & $30.40\pm0.07$ & $1.77\pm0.03$ \\
1 & VUW & Gamma & $1.0\pm0.0$ & $32.81\pm0.07$ & $1.74\pm0.03$ \\
1 & VSpS & No Reweight & \bm{$1.5508\pm0.0017$} & $21.64\pm0.06$ & $1.76\pm0.03$ \\
1 & MSE & DeltaGumbel & \bm{$1.5494\pm0.0017$} & $22.58\pm0.06$ & $1.77\pm0.03$ \\
1 & MSE & Gamma & \bm{$1.5504\pm0.0017$} & $25.15\pm0.07$ & $1.74\pm0.03$ \\
1 & MWS & DeltaGumbel & $1.5105\pm0.0017$ & $23.24\pm0.07$ & $1.77\pm0.03$ \\
1 & MWS & Gamma & $1.5312\pm0.0017$ & $25.57\pm0.07$ & $1.76\pm0.03$ \\
\midrule
2 & VSpS & No Reweight & \bm{$1.857\pm0.003$} & $19.41\pm0.07$ & $1.74\pm0.03$ \\
2 & MSE & DeltaGumbel & \bm{$1.853\pm0.003$} & $20.46\pm0.07$ & $1.78\pm0.03$ \\
2 & MSE & Gamma & \bm{$1.856\pm0.003$} & $23.42\pm0.08$ & $1.73\pm0.03$ \\
2 & MWS & DeltaGumbel & $1.773\pm0.003$ & $21.50\pm0.08$ & $1.77\pm0.03$ \\
2 & MWS & Gamma & $1.818\pm0.003$ & $25.29\pm0.09$ & $1.73\pm0.03$ \\
\midrule
3 & VSpS & No Reweight & \bm{$2.028\pm0.004$} & $19.03\pm0.08$ & $1.75\pm0.03$ \\
3 & MSE & DeltaGumbel & \bm{$2.022\pm0.004$} & $20.24\pm0.09$ & $1.77\pm0.03$ \\
3 & MSE & Gamma & \bm{$2.026\pm0.004$} & $23.98\pm0.10$ & $1.74\pm0.03$ \\
3 & MWS & DeltaGumbel & $1.913\pm0.004$ & $21.53\pm0.10$ & $1.76\pm0.03$ \\
3 & MWS & Gamma & $1.969\pm0.004$ & $25.52\pm0.11$ & $1.75\pm0.03$ \\
\midrule
4 & VSpS & No Reweight & \bm{$2.132\pm0.005$} & $19.27\pm0.09$ & $1.72\pm0.03$ \\
4 & MSE & DeltaGumbel & \bm{$2.125\pm0.005$} & $20.79\pm0.10$ & $1.73\pm0.03$ \\
4 & MSE & Gamma & \bm{$2.132\pm0.005$} & $25.28\pm0.12$ & $1.71\pm0.03$ \\
4 & MWS & DeltaGumbel & $1.987\pm0.005$ & $22.20\pm0.12$ & $1.77\pm0.03$ \\
4 & MWS & Gamma & $2.061\pm0.005$ & $27.14\pm0.14$ & $1.72\pm0.03$ \\
\bottomrule
\end{tabular}

\begin{tabular}{rllll}
\toprule
$K$ & method & reweight & ANLPPT(U Score) & ANLPPT(maximin-LLR) \\
\midrule
\midrule
1 & Basic & No Reweight & $0.0\pm0.0$ & $0.0\pm0.0$ \\
1 & VUW & DeltaGumbel & \bm{$0.376\pm0.009$} & \bm{$1.71\pm0.03$} \\
1 & VUW & Gamma & \bm{$0.097\pm0.002$} & \bm{$0.272\pm0.005$} \\
1 & VSpS & No Reweight & $0.0\pm0.0$ & $0.0\pm0.0$ \\
1 & MSE & DeltaGumbel & $0.153\pm0.004$ & $0.640\pm0.014$ \\
1 & MSE & Gamma & $0.0433\pm0.0012$ & $0.0605\pm0.0019$ \\
1 & MWS & DeltaGumbel & \bm{$0.374\pm0.009$} & \bm{$1.71\pm0.03$} \\
1 & MWS & Gamma & \bm{$0.098\pm0.002$} & \bm{$0.275\pm0.005$} \\
\midrule
2 & VSpS & No Reweight & $0.0\pm0.0$ & $0.0\pm0.0$ \\
2 & MSE & DeltaGumbel & $0.111\pm0.003$ & $0.419\pm0.010$ \\
2 & MSE & Gamma & $0.0322\pm0.0010$ & $0.0310\pm0.0014$ \\
2 & MWS & DeltaGumbel & \bm{$0.374\pm0.009$} & \bm{$1.71\pm0.03$} \\
2 & MWS & Gamma & \bm{$0.096\pm0.002$} & \bm{$0.272\pm0.005$} \\
\midrule
3 & VSpS & No Reweight & $0.0\pm0.0$ & $0.0\pm0.0$ \\
3 & MSE & DeltaGumbel & $0.094\pm0.003$ & $0.331\pm0.009$ \\
3 & MSE & Gamma & $0.0281\pm0.0009$ & $0.0214\pm0.0012$ \\
3 & MWS & DeltaGumbel & \bm{$0.374\pm0.009$} & \bm{$1.70\pm0.03$} \\
3 & MWS & Gamma & \bm{$0.097\pm0.002$} & \bm{$0.274\pm0.005$} \\
\midrule
4 & VSpS & No Reweight & $0.0\pm0.0$ & $0.0\pm0.0$ \\
4 & MSE & DeltaGumbel & $0.083\pm0.002$ & $0.280\pm0.008$ \\
4 & MSE & Gamma & $0.0258\pm0.0008$ & $0.0167\pm0.0011$ \\
4 & MWS & DeltaGumbel & \bm{$0.375\pm0.009$} & \bm{$1.71\pm0.03$} \\
4 & MWS & Gamma & \bm{$0.096\pm0.002$} & \bm{$0.271\pm0.005$} \\
\bottomrule
\end{tabular}
\caption{Text summarization task with LLaMa-7b model \cite{touvron2023llama} as target model and LLaMa-68m model \cite{miao2023specinfer} as reference model.}\label{tab:table1_summarization_scan_n_llama-7b_llama-68m}
\end{table}
 \begin{table}[H]
\centering
\begin{tabular}{rlllll}
\toprule
$K$ & method & reweight & AATPS & PTT & LOGPPL \\
\midrule
\midrule
1 & Basic & No Reweight & $1.0\pm0.0$ & $46.169\pm0.012$ & $1.27\pm0.03$ \\
1 & VUW & DeltaGumbel & $1.0\pm0.0$ & $46.956\pm0.012$ & $1.31\pm0.03$ \\
1 & VUW & Gamma & $1.0\pm0.0$ & $49.333\pm0.017$ & $1.28\pm0.03$ \\
1 & VSpS & No Reweight & \bm{$1.5651\pm0.0017$} & $31.94\pm0.06$ & $1.24\pm0.03$ \\
1 & MSE & DeltaGumbel & \bm{$1.5639\pm0.0017$} & $32.81\pm0.06$ & $1.30\pm0.03$ \\
1 & MSE & Gamma & \bm{$1.5643\pm0.0017$} & $35.35\pm0.06$ & $1.26\pm0.03$ \\
1 & MWS & DeltaGumbel & $1.5307\pm0.0017$ & $33.62\pm0.07$ & $1.31\pm0.03$ \\
1 & MWS & Gamma & $1.5476\pm0.0017$ & $35.86\pm0.07$ & $1.29\pm0.03$ \\
\midrule
2 & VSpS & No Reweight & \bm{$1.884\pm0.003$} & $27.91\pm0.09$ & $1.29\pm0.03$ \\
2 & MSE & DeltaGumbel & \bm{$1.878\pm0.003$} & $28.98\pm0.09$ & $1.34\pm0.03$ \\
2 & MSE & Gamma & \bm{$1.884\pm0.003$} & $31.89\pm0.10$ & $1.30\pm0.03$ \\
2 & MWS & DeltaGumbel & $1.815\pm0.003$ & $30.05\pm0.11$ & $1.32\pm0.03$ \\
2 & MWS & Gamma & $1.850\pm0.003$ & $33.59\pm0.11$ & $1.29\pm0.03$ \\
\midrule
3 & VSpS & No Reweight & \bm{$2.072\pm0.005$} & $26.60\pm0.11$ & $1.24\pm0.03$ \\
3 & MSE & DeltaGumbel & \bm{$2.060\pm0.005$} & $27.94\pm0.12$ & $1.31\pm0.03$ \\
3 & MSE & Gamma & \bm{$2.066\pm0.005$} & $31.63\pm0.13$ & $1.28\pm0.03$ \\
3 & MWS & DeltaGumbel & $1.972\pm0.004$ & $29.26\pm0.13$ & $1.32\pm0.03$ \\
3 & MWS & Gamma & $2.016\pm0.004$ & $33.45\pm0.15$ & $1.30\pm0.03$ \\
\midrule
4 & VSpS & No Reweight & \bm{$2.181\pm0.006$} & $26.36\pm0.12$ & $1.27\pm0.03$ \\
4 & MSE & DeltaGumbel & \bm{$2.171\pm0.006$} & $28.00\pm0.13$ & $1.30\pm0.03$ \\
4 & MSE & Gamma & \bm{$2.176\pm0.006$} & $32.44\pm0.15$ & $1.28\pm0.03$ \\
4 & MWS & DeltaGumbel & $2.059\pm0.005$ & $29.51\pm0.16$ & $1.31\pm0.03$ \\
4 & MWS & Gamma & $2.119\pm0.005$ & $34.32\pm0.17$ & $1.29\pm0.03$ \\
\bottomrule
\end{tabular}

\begin{tabular}{rllll}
\toprule
$K$ & method & reweight & ANLPPT(U Score) & ANLPPT(maximin-LLR) \\
\midrule
\midrule
1 & Basic & No Reweight & $0.0\pm0.0$ & $0.0\pm0.0$ \\
1 & VUW & DeltaGumbel & \bm{$0.260\pm0.009$} & \bm{$1.27\pm0.03$} \\
1 & VUW & Gamma & \bm{$0.067\pm0.002$} & \bm{$0.200\pm0.005$} \\
1 & VSpS & No Reweight & $0.0\pm0.0$ & $0.0\pm0.0$ \\
1 & MSE & DeltaGumbel & $0.103\pm0.004$ & $0.448\pm0.014$ \\
1 & MSE & Gamma & $0.0297\pm0.0011$ & $0.0377\pm0.0017$ \\
1 & MWS & DeltaGumbel & \bm{$0.260\pm0.009$} & \bm{$1.27\pm0.03$} \\
1 & MWS & Gamma & \bm{$0.068\pm0.002$} & \bm{$0.201\pm0.005$} \\
\midrule
2 & VSpS & No Reweight & $0.0\pm0.0$ & $0.0\pm0.0$ \\
2 & MSE & DeltaGumbel & $0.076\pm0.003$ & $0.292\pm0.010$ \\
2 & MSE & Gamma & $0.0226\pm0.0009$ & $0.0174\pm0.0012$ \\
2 & MWS & DeltaGumbel & \bm{$0.261\pm0.009$} & \bm{$1.27\pm0.03$} \\
2 & MWS & Gamma & \bm{$0.069\pm0.002$} & \bm{$0.201\pm0.005$} \\
\midrule
3 & VSpS & No Reweight & $0.0\pm0.0$ & $0.0\pm0.0$ \\
3 & MSE & DeltaGumbel & $0.063\pm0.002$ & $0.224\pm0.008$ \\
3 & MSE & Gamma & $0.0199\pm0.0008$ & $0.0098\pm0.0011$ \\
3 & MWS & DeltaGumbel & \bm{$0.261\pm0.009$} & \bm{$1.27\pm0.03$} \\
3 & MWS & Gamma & \bm{$0.068\pm0.002$} & \bm{$0.201\pm0.006$} \\
\midrule
4 & VSpS & No Reweight & $0.0\pm0.0$ & $0.0\pm0.0$ \\
4 & MSE & DeltaGumbel & $0.057\pm0.002$ & $0.192\pm0.007$ \\
4 & MSE & Gamma & $0.0184\pm0.0007$ & $0.0069\pm0.0009$ \\
4 & MWS & DeltaGumbel & \bm{$0.260\pm0.009$} & \bm{$1.27\pm0.03$} \\
4 & MWS & Gamma & \bm{$0.068\pm0.002$} & \bm{$0.202\pm0.005$} \\
\bottomrule
\end{tabular}
\caption{Text summarization task with LLaMa-13b model \cite{touvron2023llama} as target model and LLaMa-68m model \cite{miao2023specinfer} as reference model.}\label{tab:table1_summarization_scan_n_llama-13b_llama-68m}
\end{table}
 \begin{table}[H]
\centering
\begin{tabular}{rlllll}
\toprule
$K$ & method & reweight & AATPS & PTT & LOGPPL \\
\midrule
\midrule
1 & Basic & No Reweight & $1.0\pm0.0$ & $30.20\pm0.08$ & $2.161\pm0.014$ \\
1 & VUW & DeltaGumbel & $1.0\pm0.0$ & $31.06\pm0.08$ & $2.147\pm0.014$ \\
1 & VUW & Gamma & $1.0\pm0.0$ & $33.34\pm0.09$ & $2.158\pm0.014$ \\
1 & VSpS & No Reweight & \bm{$1.5004\pm0.0016$} & $22.69\pm0.07$ & $2.158\pm0.014$ \\
1 & MSE & DeltaGumbel & \bm{$1.5004\pm0.0016$} & $23.52\pm0.07$ & $2.150\pm0.014$ \\
1 & MSE & Gamma & \bm{$1.5001\pm0.0016$} & $26.30\pm0.07$ & $2.160\pm0.014$ \\
1 & MWS & DeltaGumbel & $1.4546\pm0.0016$ & $24.63\pm0.07$ & $2.147\pm0.014$ \\
1 & MWS & Gamma & $1.4753\pm0.0016$ & $26.67\pm0.08$ & $2.151\pm0.014$ \\
\midrule
2 & VSpS & No Reweight & \bm{$1.755\pm0.003$} & $21.00\pm0.07$ & $2.145\pm0.014$ \\
2 & MSE & DeltaGumbel & \bm{$1.753\pm0.003$} & $21.93\pm0.07$ & $2.152\pm0.014$ \\
2 & MSE & Gamma & \bm{$1.754\pm0.003$} & $25.00\pm0.08$ & $2.159\pm0.014$ \\
2 & MWS & DeltaGumbel & $1.665\pm0.003$ & $23.11\pm0.07$ & $2.147\pm0.014$ \\
2 & MWS & Gamma & $1.704\pm0.003$ & $27.28\pm0.09$ & $2.145\pm0.014$ \\
\midrule
3 & VSpS & No Reweight & \bm{$1.883\pm0.004$} & $20.57\pm0.08$ & $2.153\pm0.014$ \\
3 & MSE & DeltaGumbel & \bm{$1.881\pm0.004$} & $22.08\pm0.08$ & $2.148\pm0.014$ \\
3 & MSE & Gamma & \bm{$1.882\pm0.004$} & $25.97\pm0.09$ & $2.143\pm0.014$ \\
3 & MWS & DeltaGumbel & $1.763\pm0.004$ & $23.71\pm0.08$ & $2.145\pm0.014$ \\
3 & MWS & Gamma & $1.817\pm0.004$ & $28.23\pm0.10$ & $2.153\pm0.014$ \\
\midrule
4 & VSpS & No Reweight & \bm{$1.950\pm0.005$} & $21.35\pm0.08$ & $2.153\pm0.014$ \\
4 & MSE & DeltaGumbel & \bm{$1.946\pm0.005$} & $23.10\pm0.09$ & $2.153\pm0.014$ \\
4 & MSE & Gamma & \bm{$1.951\pm0.005$} & $28.10\pm0.11$ & $2.143\pm0.014$ \\
4 & MWS & DeltaGumbel & $1.809\pm0.004$ & $24.65\pm0.09$ & $2.146\pm0.014$ \\
4 & MWS & Gamma & $1.872\pm0.004$ & $30.24\pm0.12$ & $2.148\pm0.014$ \\
\bottomrule
\end{tabular}

\begin{tabular}{rllll}
\toprule
$K$ & method & reweight & ANLPPT(U Score) & ANLPPT(maximin-LLR) \\
\midrule
\midrule
1 & Basic & No Reweight & $0.0\pm0.0$ & $0.0\pm0.0$ \\
1 & VUW & DeltaGumbel & \bm{$0.467\pm0.005$} & \bm{$2.101\pm0.014$} \\
1 & VUW & Gamma & \bm{$0.1291\pm0.0015$} & \bm{$0.3762\pm0.0018$} \\
1 & VSpS & No Reweight & $0.0\pm0.0$ & $0.0\pm0.0$ \\
1 & MSE & DeltaGumbel & $0.189\pm0.003$ & $0.771\pm0.008$ \\
1 & MSE & Gamma & $0.0567\pm0.0010$ & $0.0862\pm0.0014$ \\
1 & MWS & DeltaGumbel & \bm{$0.466\pm0.005$} & \bm{$2.100\pm0.014$} \\
1 & MWS & Gamma & \bm{$0.1283\pm0.0015$} & \bm{$0.3759\pm0.0018$} \\
\midrule
2 & VSpS & No Reweight & $0.0\pm0.0$ & $0.0\pm0.0$ \\
2 & MSE & DeltaGumbel & $0.133\pm0.002$ & $0.494\pm0.007$ \\
2 & MSE & Gamma & $0.0411\pm0.0008$ & $0.0455\pm0.0011$ \\
2 & MWS & DeltaGumbel & \bm{$0.467\pm0.005$} & \bm{$2.101\pm0.014$} \\
2 & MWS & Gamma & \bm{$0.1285\pm0.0015$} & \bm{$0.3752\pm0.0017$} \\
\midrule
3 & VSpS & No Reweight & $0.0\pm0.0$ & $0.0\pm0.0$ \\
3 & MSE & DeltaGumbel & $0.1116\pm0.0019$ & $0.385\pm0.006$ \\
3 & MSE & Gamma & $0.0354\pm0.0008$ & $0.0309\pm0.0010$ \\
3 & MWS & DeltaGumbel & \bm{$0.466\pm0.005$} & \bm{$2.099\pm0.014$} \\
3 & MWS & Gamma & \bm{$0.1292\pm0.0015$} & \bm{$0.3765\pm0.0017$} \\
\midrule
4 & VSpS & No Reweight & $0.0\pm0.0$ & $0.0\pm0.0$ \\
4 & MSE & DeltaGumbel & $0.1030\pm0.0018$ & $0.340\pm0.005$ \\
4 & MSE & Gamma & $0.0325\pm0.0007$ & $0.0250\pm0.0009$ \\
4 & MWS & DeltaGumbel & \bm{$0.465\pm0.005$} & \bm{$2.099\pm0.014$} \\
4 & MWS & Gamma & \bm{$0.1286\pm0.0015$} & \bm{$0.3761\pm0.0017$} \\
\bottomrule
\end{tabular}
\caption{Open-ended text generation task with LLaMa-7b model \cite{touvron2023llama} as target model and LLaMa-68m model \cite{miao2023specinfer} as reference model.}\label{tab:table1_oeg_scan_n_llama-7b_llama-68m}
\end{table}
 \begin{table}[H]
\centering
\begin{tabular}{rlllll}
\toprule
$K$ & method & reweight & AATPS & PTT & LOGPPL \\
\midrule
\midrule
1 & Basic & No Reweight & $1.0\pm0.0$ & $44.982\pm0.012$ & $2.087\pm0.014$ \\
1 & VUW & DeltaGumbel & $1.0\pm0.0$ & $45.766\pm0.012$ & $2.062\pm0.013$ \\
1 & VUW & Gamma & $1.0\pm0.0$ & $48.030\pm0.017$ & $2.086\pm0.014$ \\
1 & VSpS & No Reweight & \bm{$1.4879\pm0.0016$} & $32.55\pm0.04$ & $2.076\pm0.013$ \\
1 & MSE & DeltaGumbel & \bm{$1.4891\pm0.0016$} & $33.40\pm0.04$ & $2.071\pm0.013$ \\
1 & MSE & Gamma & \bm{$1.4883\pm0.0016$} & $35.78\pm0.05$ & $2.073\pm0.014$ \\
1 & MWS & DeltaGumbel & $1.4439\pm0.0016$ & $34.52\pm0.05$ & $2.063\pm0.013$ \\
1 & MWS & Gamma & $1.4636\pm0.0016$ & $36.55\pm0.05$ & $2.073\pm0.013$ \\
\midrule
2 & VSpS & No Reweight & \bm{$1.734\pm0.003$} & $29.32\pm0.06$ & $2.071\pm0.013$ \\
2 & MSE & DeltaGumbel & \bm{$1.732\pm0.003$} & $30.39\pm0.06$ & $2.074\pm0.014$ \\
2 & MSE & Gamma & \bm{$1.731\pm0.003$} & $33.50\pm0.07$ & $2.081\pm0.014$ \\
2 & MWS & DeltaGumbel & $1.646\pm0.003$ & $32.07\pm0.07$ & $2.066\pm0.013$ \\
2 & MWS & Gamma & $1.683\pm0.003$ & $35.64\pm0.07$ & $2.076\pm0.013$ \\
\midrule
3 & VSpS & No Reweight & \bm{$1.854\pm0.004$} & $28.76\pm0.07$ & $2.071\pm0.014$ \\
3 & MSE & DeltaGumbel & \bm{$1.853\pm0.004$} & $30.05\pm0.08$ & $2.075\pm0.014$ \\
3 & MSE & Gamma & \bm{$1.855\pm0.004$} & $33.97\pm0.09$ & $2.070\pm0.013$ \\
3 & MWS & DeltaGumbel & $1.738\pm0.004$ & $32.15\pm0.08$ & $2.064\pm0.013$ \\
3 & MWS & Gamma & $1.790\pm0.004$ & $36.37\pm0.09$ & $2.080\pm0.014$ \\
\midrule
4 & VSpS & No Reweight & \bm{$1.913\pm0.004$} & $29.05\pm0.08$ & $2.074\pm0.014$ \\
4 & MSE & DeltaGumbel & \bm{$1.914\pm0.004$} & $30.69\pm0.09$ & $2.072\pm0.013$ \\
4 & MSE & Gamma & \bm{$1.915\pm0.004$} & $35.47\pm0.10$ & $2.075\pm0.014$ \\
4 & MWS & DeltaGumbel & $1.781\pm0.004$ & $33.09\pm0.09$ & $2.062\pm0.013$ \\
4 & MWS & Gamma & $1.836\pm0.004$ & $38.21\pm0.11$ & $2.072\pm0.013$ \\
\bottomrule
\end{tabular}

\begin{tabular}{rllll}
\toprule
$K$ & method & reweight & ANLPPT(U Score) & ANLPPT(maximin-LLR) \\
\midrule
\midrule
1 & Basic & No Reweight & $0.0\pm0.0$ & $0.0\pm0.0$ \\
1 & VUW & DeltaGumbel & \bm{$0.441\pm0.005$} & \bm{$2.016\pm0.013$} \\
1 & VUW & Gamma & \bm{$0.1245\pm0.0015$} & \bm{$0.3692\pm0.0017$} \\
1 & VSpS & No Reweight & $0.0\pm0.0$ & $0.0\pm0.0$ \\
1 & MSE & DeltaGumbel & $0.176\pm0.003$ & $0.724\pm0.008$ \\
1 & MSE & Gamma & $0.0527\pm0.0009$ & $0.0794\pm0.0013$ \\
1 & MWS & DeltaGumbel & \bm{$0.441\pm0.005$} & \bm{$2.017\pm0.013$} \\
1 & MWS & Gamma & \bm{$0.1226\pm0.0014$} & \bm{$0.3680\pm0.0017$} \\
\midrule
2 & VSpS & No Reweight & $0.0\pm0.0$ & $0.0\pm0.0$ \\
2 & MSE & DeltaGumbel & $0.122\pm0.002$ & $0.457\pm0.006$ \\
2 & MSE & Gamma & $0.0387\pm0.0008$ & $0.0405\pm0.0011$ \\
2 & MWS & DeltaGumbel & \bm{$0.442\pm0.005$} & \bm{$2.019\pm0.013$} \\
2 & MWS & Gamma & \bm{$0.1238\pm0.0014$} & \bm{$0.3691\pm0.0017$} \\
\midrule
3 & VSpS & No Reweight & $0.0\pm0.0$ & $0.0\pm0.0$ \\
3 & MSE & DeltaGumbel & $0.1036\pm0.0018$ & $0.357\pm0.006$ \\
3 & MSE & Gamma & $0.0330\pm0.0007$ & $0.0279\pm0.0010$ \\
3 & MWS & DeltaGumbel & \bm{$0.442\pm0.005$} & \bm{$2.018\pm0.013$} \\
3 & MWS & Gamma & \bm{$0.1240\pm0.0014$} & \bm{$0.3691\pm0.0017$} \\
\midrule
4 & VSpS & No Reweight & $0.0\pm0.0$ & $0.0\pm0.0$ \\
4 & MSE & DeltaGumbel & $0.0943\pm0.0017$ & $0.311\pm0.005$ \\
4 & MSE & Gamma & $0.0308\pm0.0007$ & $0.0219\pm0.0009$ \\
4 & MWS & DeltaGumbel & \bm{$0.441\pm0.005$} & \bm{$2.016\pm0.013$} \\
4 & MWS & Gamma & \bm{$0.1237\pm0.0014$} & \bm{$0.3689\pm0.0017$} \\
\bottomrule
\end{tabular}
\caption{Open-ended text generation task with LLaMa-13b model \cite{touvron2023llama} as target model and LLaMa-68m model \cite{miao2023specinfer} as reference model.}\label{tab:table1_oeg_scan_n_llama-13b_llama-68m}
\end{table}

\newpage

\end{document}